\documentclass[a4paper, 12pt]{article}
\usepackage{subfig}
\usepackage{float}
\usepackage{wrapfig}
\usepackage{graphicx}
\usepackage{amsmath}
\usepackage{amssymb}
\usepackage{booktabs}
\usepackage[top=3.5cm, bottom=2.5cm, left=3.5cm, right=3.5cm]{geometry}

\newcommand{\hb}{H$\beta$}
\newcommand{\ha}{H$\alpha$}
\newcommand{\oiii}{[OIII]}
\newcommand{\oii}{[OII]}
\newcommand{\nii}{[NII]}

\newcommand{\esc}{erg cm$^{-2}$ s$^{-1}$}
\newcommand{\es}{erg s$^{-1}$}

\def\simgt{\lower.5ex\hbox{\gtsima}}
\def\simlt{\lower.5ex\hbox{\ltsima}}

\begin{document}
\pagestyle{empty}
\noindent
\textbf{An extremely young massive clump forming by gravitational collapse in a primordial galaxy} 
\\
\\
A. Zanella$^{1}$, E. Daddi$^1$, E. Le Floc'h$^1$, F. Bournaud$^1$, R. Gobat$^{1,2}$,
F. Valentino$^1$, V. Strazzullo$^{1,3}$, A. Cibinel$^{1,4}$, M. Onodera$^5$, V. Perret$^6$,
F. Renaud$^{1,7}$, C. Vignali$^{8,9}$
\\
\\
\scriptsize
{$^1$ Laboratoire AIM-Paris-Saclay, CEA/DSM-CNRS-Universit\'e Paris
Diderot, Irfu/Service d'Astrophysique, CEA Saclay, Orme des Merisiers,
F-91191 Gif sur Yvette, France. Email: anita.zanella@cea.fr
\\
$^2$ School of Physics, Korea Institute for Advanced Study, Heogiro
85, Seoul 130-722, Republic of Korea
\\
$^3$ Department of Physics, Ludwig-Maximilians-Universit\"at,
Scheinerstr. 1, 81679 M\"unchen, Germany
\\
$^4$ Astronomy Centre, Department of Physics and Astronomy,
University of Sussex, Brighton, BN1 9QH, UK
\\
$^5$ Institute for Astronomy, ETH Z\"urich, Wolfgang-Pauli-strasse 27, 8093 Z\"urich, Switzerland
\\
$^6$ Aix Marseille Universit\'e, CNRS, LAM (Laboratoire d’Astrophysique de Marseille), F-13388 Marseille, France
\\
$^7$ Department of Physics, University of Surrey, Guildford GU2 7XH, UK
\\
$^8$ Dipartimento di Fisica e Astronomia, Universit\`a degli Studi di
Bologna, Viale Berti Pichat 6/2, 40127 Bologna, Italy
\\
$^9$ INAF – Osservatorio Astronomico di Bologna, Via Ranzani 1,
40127 Bologna, Italy
\normalsize
\\
\\
\textbf{When the cosmic star formation history peaks ($\mathrm{\mathbf{z\sim2}}$), galaxies vigorously fed by cosmic reservoirs$^{1,2}$ are gas dominated$^{3,4}$ and contain massive star-forming clumps$^{5,6}$, thought to form by violent  gravitational instabilities in highly turbulent gas-rich disks$^{7,8}$. However, a clump formation event has not been witnessed yet, and it is debated whether clumps survive energetic feedback from young stars, thus migrating inwards to form galaxy bulges$^{9,10,11,12}$. Here we report spatially resolved spectroscopy of a bright off-nuclear emission line region in a galaxy at $\mathrm{\mathbf{z=1.987}}$. Although this region dominates the star formation in the galaxy disk, its stellar continuum remains undetected in deep imaging, revealing an extremely young (age $<$ 10 Myr) massive clump, forming through the gravitational collapse of  $> \mathrm{\mathbf{10^9\, M_\odot}}$ of gas. Gas consumption in this young clump is $\mathbf{>\,10\times}$ faster than in the host galaxy, displaying high star formation efficiency during this phase, in agreement with our hydrodynamic simulations.
The frequency of older clumps with similar masses$^{13}$ coupled with our initial estimate of their formation rate ($\mathbf{\sim2.5}$~Gyr$^{\mathbf{-1}}$) supports long lifetimes  ($\mathbf{\sim500}$~Myr), favouring scenarios where clumps survive feedback and grow the bulges of present-day galaxies.}
\\
\\
The high spatial resolution and sensitivity of Hubble Space Telescope (\textit{HST}) imaging and spectroscopy routinely allows us to resolve giant star-forming regions (clumps) inside galaxies at $\mathrm{z} \sim 2$, three billion years after the Big Bang. Stellar population modelling has revealed a wide range of ages for clumps observed in the continuum$^{6,14,15,16}$, with average age $\sim100$~Myr.  Yet clump formation rates and lifetimes remain poorly constrained$^{11,14,17,18}$. Continuum-based stellar ages are likely underestimated since clumps lose stars and re-accrete gas during their evolution$^{9}$, while very young ages  ($<30$~Myr) cannot be probed with continuum imaging alone. High equivalent width (EW) emission lines are required. 

We obtained 16 orbits of  \textit{HST} Wide Field Camera~3 G141 slitless spectroscopy and imaging with the F140W, F105W and F606W filters targeting a galaxy cluster at $\mathrm{z} = 2\,^{19}$. The F606W band traces the star formation distribution in the UV rest-frame, while the F140W probes the optical rest-frame, reflecting the stellar mass distribution.  Nebular [OIII]$\lambda$5007\AA\ emission was detected for 68 galaxies with stellar masses $9.5 < \mathrm{log(M/M_{\odot})} < 11.5$ and  redshift $1.3 \leq  \mathrm{z_{spec}} \leq 2.3$, with measurements or upper limits  for \hb, [OII]$\lambda$3727\AA ~and \ha ~when available. From spatially resolved emission line maps  we discovered a galaxy at $\mathrm{z} = 1.987$ with a remarkably bright, off-nuclear emission line region ($\mathrm{F_{\oiii}} = 4.3\pm 0.2 \times  10^{-17}$ \esc,  observed, plus \hb ~and \oii; Methods), lacking any obvious counterpart in broad-band imaging (Figure \ref{fig:maps}). The \oiii ~emission is spatially unresolved (radius $<$\,500\,pc) and located at the apparent distance of $1.6 \pm 0.3$ kpc (offset significance $7.6\sigma$, Methods) from the nucleus  (i.e., the barycenter of the stellar continuum). The deprojected distance is constrained within $3.6 \leq \mathrm{d} \leq 6.2$~kpc, corresponding to 1.3 -- 2.2 times the galaxy half-light radius (Methods). Subtracting a point-like emission leaves no significant residuals in the \oiii\ map. The continuum reddening and mass-to-light ratio (M/L) maps are flat over the galaxy, excluding that the feature is artificially induced by dust lanes or inhomogeneous attenuation
(Extended Data [ED] Figure 1 -- 2; Methods).

From emission line ratios we estimated a reddening $\mathrm{E(B-V)\sim0.3}$ and a gas-phase metallicity $\mathrm{Z\sim0.4\, Z_\odot}$ for this region, consistent with the host galaxy. Robust upper limits on its stellar continuum were estimated with detailed simulations, leading to remarkably high emission line EWs lower limits.
Given these limits and the line luminosities, the emitting region cannot be powered by a massive black hole nor by shock ionisation from wind outflows. We similarly disfavor the hypothesis of a transient, since the line luminosities remain constant over time, or an {\it ex situ\,} merging system, since its older underlying stellar continuum would be detected$^{20}$.
Also, this galaxy is classified as a disk (not a merger) from its Asymmetry and M$_{20}$ parameters (Methods). 
Therefore, a young star-forming clump  formed {\it in situ} is the most plausible interpretation.  On the basis of stellar population synthesis modelling for galaxies with active star formation, the observed EWs require very young ages for the star formation event, with a firm upper limit of 10\,Myr (Figure \ref{fig:ew}). Thus, while the ubiquity of clumps in high-z galaxies has been known for a decade, we are witnessing here for the first time the formation of a star-forming clump in the early stage of its gravitational collapse.
From the reddening corrected line luminosities   we estimate a clump star formation rate SFR $=32\pm 6 ~ \mathrm{M_{\odot}\, yr^{-1}}$, comparable to the rest of the host galaxy disk (Methods). The F140W continuum non-detection translates into a stellar mass limit   $\mathrm{M_{\star} \lesssim 3\cdot 10^8~M_{\odot}}$. To infer the underlying gas mass of the clump we considered the Jeans mass of the galaxy as a plausible upper limit, as fragmentation at higher masses is unlikely. This constrains the clump gas mass to $\mathrm{M_{gas} \lesssim 2.5 \cdot 10^9 ~M_{\odot}}$, assuming a maximal gas velocity dispersion $\sigma_V \sim 80$ km s$^{-1\, 21,22}$ (Methods).

This finding offers new insights into the physics of {\em clump formation} in gas-rich turbulent media at high-z. 
Using the estimate of its underlying gas mass, stellar mass and SFR we can constrain the nature of its star formation mode. 
Its specific star formation rate ($\mathrm{sSFR = SFR/M_{\star}}$) is $>30$ times higher than that of its host galaxy, a typical Main Sequence (MS) galaxy at z $\sim$ 2. Similarly, the lower limit on the clump star formation efficiency ($\mathrm{SFE = SFR/M_{gas}}$) is $>10$ times higher than that of normal galaxies (Figure \ref{fig:sk}), a behaviour that at galaxy-wide scales is only observed for extreme starbursts$^{23}$. At sub-galactic scales such a high SFE is observed for nearby molecular clouds$^{24}$, which are small and transient features a thousand times less massive than the present clump. Possibly at odds with what has been assumed so far$^{13, 17, 18}$, this provides observational evidence that giant clumps do not follow the Schmidt-Kennicutt law of normal star-forming galaxies, at least in the early stages of collapse. Instead, this luminous sub-galactic structure appears to follow the universal star formation law  normalized by the dynamical time$^{25,26,27}$. Comparing with the SFRs reported for older clumps with similar masses$^{13}$, we estimated a SFR enhancement of $\sim$ 3 -- 5 at ``peak formation'' with respect to later phases. 
This is the first observation of a massive star-forming clump with a robust stellar age estimate that is similar to or shorter than its dynamical time (Methods). 

Prompted by our observations, we investigated the properties of clumps in their formation phase using high resolution simulations$^{9}$. We solved the dark matter, stellar and gas gravity and hydrodynamics at a resolution of 3.5\,pc, gas cooling down to 100\,K, and we modelled the feedback processes from young stars onto the gas: photo-ionization, radiation pressure and supernovae explosions. Figure~\ref{fig:simulations} shows a typical, $\mathrm{M_{\star}} \sim 3 \times 10^{10}\,\mathrm{M_{\odot}}$, $\mathrm{z}=2$ galaxy model with giant clumps formed through violent disk instability. Their formation sites are located at 2.1 -- 7.0\,kpc from the nucleus of the galaxy that has an half-mass radius of 4.5\,kpc, consistent with many other simulations$^{10, 20, 28}$ and our observations. All clumps, and especially the youngest ones, are brighter in the SFR map than in the continuum: they undergo a burst of star formation during their initial collapse, with peak SFRs about $10\, -\, 20~\mathrm{M_{\odot}\, yr^{-1}}$ consistent with our observations, then evolve to a lower sSFR regime within 20\,Myr, once feedback regulates star formation and their stellar mass has grown. Our simulations further corroborate the idea that all clumps behave like galactic miniatures of starbursts  in the Schmidt-Kennicutt diagram during their first 20\,Myr (Figure \ref{fig:sk}). 
The SFE of simulated clumps decreases at later times, although it remains $\gtrsim$ 0.5\,dex higher than MS galaxies, consistent with their shorter dynamical times. The presence of  massive clumps is probably an effective reason for the observed rise of the SFE in normal MS galaxies  from z = 0 to z = 2$^{3,4,29}$, given the increasing prevalence of clumps at high-z. 
Furthermore, the violent burst-like behaviour that young clumps show at formation is consistent with simulations predicting that, thanks to their rapid collapse, giant clumps could form globular clusters by converting gas into stars faster than stars expel the gas$^{30}$.

The  short visibility window at high EWs ($\lesssim10$~Myr independently of the star formation history) has likely prevented until now the detection of the clumps formation phase. From this timing constraint and the single discovery in our survey, we attempted a first estimate of the clump formation rate of 2.5~Gyr$^{-1}$ per galaxy (for $\mathrm{M_{\rm clump} \gtrsim 2.5\times10^9\, M_\odot}$, Methods). 
Given the observation of 1 -- 2~clumps per galaxy with similar masses$^{13,17,18}$, this converts into a  lifetime of $\approx$ 500\,Myr (Methods). This is longer than expected in models of clump destruction by stellar feedback$^{10,28}$. Instead, it is representative of the timescale needed for giant clumps formed in galactic disks to migrate inward through dynamical friction and gravity torques and coalesce to grow the central galactic bulges$^{9,12}$.

Our study demonstrates the detectability of ultra-young clumps in deep surveys, indicating low formation rates and long lifetimes. This is crucial to understand key issues of galaxy formation and evolution such as clumps migration, bulge formation and the role of feedback. However, future observations of larger samples of forming clumps with direct measurements of clumps' sizes, gas masses and velocity widths (and hence dynamical masses) are required for a definitive understanding. This should be within the capabilities of the complete Atacama Large Millimeter Array and James Webb Space Telescope.
We note that spectroscopic surveys targeting high-z galaxies (e.g. SINS, 3D-HST) have not yet reported the identification of giant  clumps at formation. This might suggest that they are rarer events than what appears from our survey, which finally allowed us to identify a direct signature of massive clump formation via gravitational collapse.

\newpage
\textbf{References}
\\
1. Kere\v s, D. How do galaxies get their gas? \textit{Mon. Notic. R. Astron. Soc.} \textbf{363} 2 - 28 (2005)
\\
2. Dekel, A. et al. Cold streams in early massive hot halos as the main mode of galaxy formation. \textit{Nature} \textbf{457} 451 - 454 (2009)
\\
3. Daddi, E. et al. Very high gas fractions and extended gas
reservoirs in z = 1.5 disk galaxies. \textit{Astroph. J.} \textbf{713} 686-707 (2010)
\\
4. Tacconi, L. J. et al. High molecular gas fractions in normal
massive star forming galaxies in the young Universe \textit{Nature} \textbf{463} 781-784 (2010)
\\
5. Genzel, R. et al. The rapid formation of a large rotating disk
galaxy three billion years after the Big Bang. \textit{Nature} \textbf{442} 786-789 (2006)
\\
6. Elmegreen, B. G. et al. Bulge and clump evolution in Hubble Ultra Deep
Field clump clusters, chains and spiral galaxies. \textit{Astroph. J.} \textbf{692} 12-31 (2009)
\\
7. Elmegreen, D. M., Elmegreen, B. G., \& Hirst, A. C. Discovery of Face-on Counterparts of Chain Galaxies in the Tadpole Advanced Camera for Surveys Field. \textit{Astrophys. J. Lett.} \textbf{604} 21 - 23 (2004)
\\
8. Bournaud, F., Elmegreen B. G. \& Elmegreen, D. M. Rapid Formation of Exponential Disks and Bulges at High Redshift from the Dynamical Evolution of Clump-Cluster and Chain Galaxies. \textit{Astrophys. J.} \textbf{670} 237 - 248 (2007)
\\
9. Bournaud, F. et al. The long lives of giant clumps and the birth of
outflows in gas-rich galaxies at high-redshift. \textit{Astroph. J.} \textbf{780} 57-75 (2014)
\\
10. Genel, S. et al. Short-lived star-forming giant clumps in
cosmological simulations of z $\sim$ 2 disks. \textit{Astroph. J.} \textbf{745} 11-21 (2012)
\\
11. Wuyts, S. et al. A CANDELS-3D-HST synergy: resolved star formation
patterns at 0.7 $< \mathrm{z} < 1.5$. \textit{Astroph. J.} \textbf{779} 135-151 (2013)
\\
12. Dekel, A. et al. Formation of massive galaxies at high redshift:
cold streams, clumpy disks, and compact spheroids. \textit{Astroph. J.} \textbf{703} 785-801 (2009)
\\
13. F\"orster-Schreiber, N. et al. Constraints on the assembly and
dynamics of galaxies. II. Properties of kiloparsec-scale clumps in
rest-frame optical emission of z $\sim$ 2 star-forming
galaxies. \textit{Astroph. J.} \textbf{739} 45-69 (2011)
\\
14. Wuyts, S. et al. Smooth(er) stellar mass maps in CANDELS:
constraints on the longevity of clumps in high-redshift star-forming
galaxies. \textit{Astrophys. J.} \textbf{753} 114-139 (2012)
\\
15. Guo, Y. et al. Multi-wavelength view of kiloparsec-scale clumps in
star-forming galaxies at z $\sim$ 2. \textit{Astrophys. J.}
\textbf{757} 120-142 (2012)
\\
16. Elmegreen, B. et al. Massive clumps in local galaxies: comparison
with high-redshift clumps. \textit{Astrophys. J.} \textbf{774} 86-100 (2013)
\\
17. Genzel, R. et al. The Sins survey of z $\sim$ 2 alaxy kinematics:
properties of the giant star-forming clumps. \textit{Astrophys. J.}
\textbf{733} 101-131 (2011)
\\
18. Newman, S. et al. The SINS/zC-SINF survey of z $\sim$ 2 galaxy
kinematics: outflow properties. \textit{Astrophys. J.} \textbf{761} 43-50 (2012)
\\
19. Gobat, R. et al. WFC3 GRISM comfirmation of the distant cluster CL
J1449+0856 at $\langle \mathrm{z} \rangle = 2.00$: quiescent and star-forming
galaxy populations \textit{Astrophys. J.} \textbf{776} 9-21 (2013)
\\
20. Mandelker, N. et al. The population of giant clumps in simulated
high-z galaxies: in situ and ex situ migration and
survival. \textit{Mon. Notic. R. Astron. Soc.} \textbf{443} 3675 -
3702 (2014)
\\
21. F\"orster-Schreiber, N. et al. The SINS survey: Sinfoni integral
field spectroscopy of z $\sim$ 2 star-forming
galaxies. \textit{Astrophys. J.} \textbf{706} 1364-1428 (2009)
\\
22. Contini, T. et al. MASSIV: Mass Assembly Survey with Sinfoni in
VVDS. \textit{Astron. \& Astrophys.} \textbf{593} 91 (2012)
\\
23. Rodighiero, G. et al. The Lesser Role of Starbursts in Star
Formation at z = 2. \textit{Astroph. J. Lett.} \textbf{739} 40 (2011)
\\
24. Lada, C.~J., Lombardi, M. and Alves, J.~F. On the Star Formation Rates in Molecular Clouds. \textit{Astroph. J.} \textbf{724} 687-693 (2010)
\\
25. Daddi, E. et al. Different Star Formation Laws for Disks Versus
Starbursts at Low and High Redshifts. \textit{Astroph. J. Lett.}
\textbf{714} 118 - 122 (2010)
\\
26. Genzel, R. et al. A study of the gas-star formation relation over
cosmic time \textit{Mon. Notic. R. Astron. Soc.} \textbf{407} 2091 -
2108 (2010)
\\
27. Krumholz, M. R., Dekel, A. \& McKee, C. F. A Universal, Local Star
Formation Law in Galactic Clouds, nearby Galaxies, High-redshift
Disks, and Starbursts \textit{Astroph. J.} \textbf{645} 69 (2012)
\\
28. Hopkins, P. F. et al. Stellar feedback and bulge formation in clumpy discs \textit{Mont. Notic. R. Astron. Soc.} \textbf{427} 968 - 978 (2012)
\\
29. Sargent, M. T. et al. Regularity underlying complexity: a
redshift-independent description of the continuous variation of
galaxy-scale molecular gas properties in the mass-star formation rate
plane. \textit{Astroph. J.} \textbf{793} 19 (2014)
\\
30. Shapiro, K. L., Genzel, R. and F\"orster Schreiber,
N. M. Star-forming galaxies at z $\sim$ 2 and the formation of the
metal-rich globular cluster population
\textit{Mon. Notic. R. Astron. Soc.} \textbf{403} 36-40 (2010)
\\
\\
\textbf{Acknowledgements} We thank Stephanie Juneau for
discussions and Michele Cappellari for sharing his Multi-Gaussian
Expansion fit software publicly. The simulations were performed at the {\em Tr\'es Grand Centre de Calcul} of CEA under GENCI allocation 2014-GEN2192.  
We acknowledge financial support from Agence Nationale de la Recherche
(contract \#ANR-12-JS05-0008-01) and EC through an ERC grants
StG-257720 and StG-240039.
\\
\\
\textbf{Author Contributions} All authors have extensively contributed
to this work.
\\
\\
\textbf{Author Information} Reprints and permissions information is
available at www.nature.com/reprints. The authors have no competing
financial interests. Correspondence and requests for materials should
be addressed to A. Zanella (anita.zanella@cea.fr).

\newpage
\begin{figure}[h!]
\centering
\includegraphics[width=\textwidth]{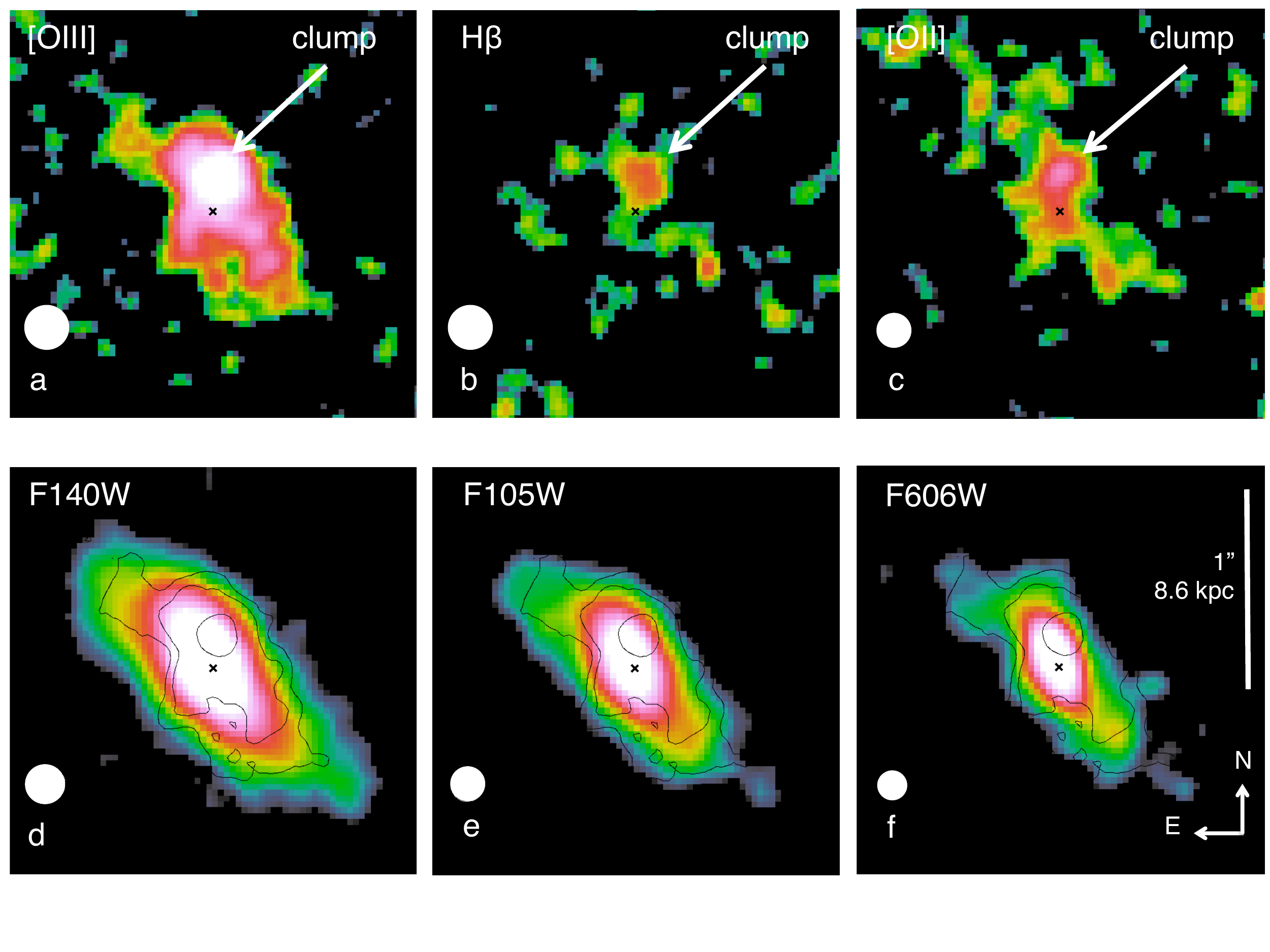}
\caption{A massive, very young clump in a disk galaxy at $\mathrm{z = 1.987}$. 
The emission line maps show off-nuclear, unresolved, bright [OIII] together with H$\beta$ and [OII] emissions (\textbf{a}
- \textbf{c}), respectively 18$\sigma$, 3$\sigma$ and 3.5$\sigma$ significant.
No counterpart in the direct images is detected (\textbf{d} - \textbf{f}). The flux contours of the \oiii ~map have been
overplotted on the direct images. The color scales logarithmically with
flux from the minimum (black) to maximum (white) level displayed (different
for \textbf{a} - \textbf{c}, \textbf{d} - \textbf{f}). The black cross in each panel indicates
the barycenter of the stellar optical rest-frame continuum and the white circle the PSF FWHM.}
\label{fig:maps}
\end{figure}

\newpage

\begin{figure}[h!]
\centering
\includegraphics[width=\textwidth]{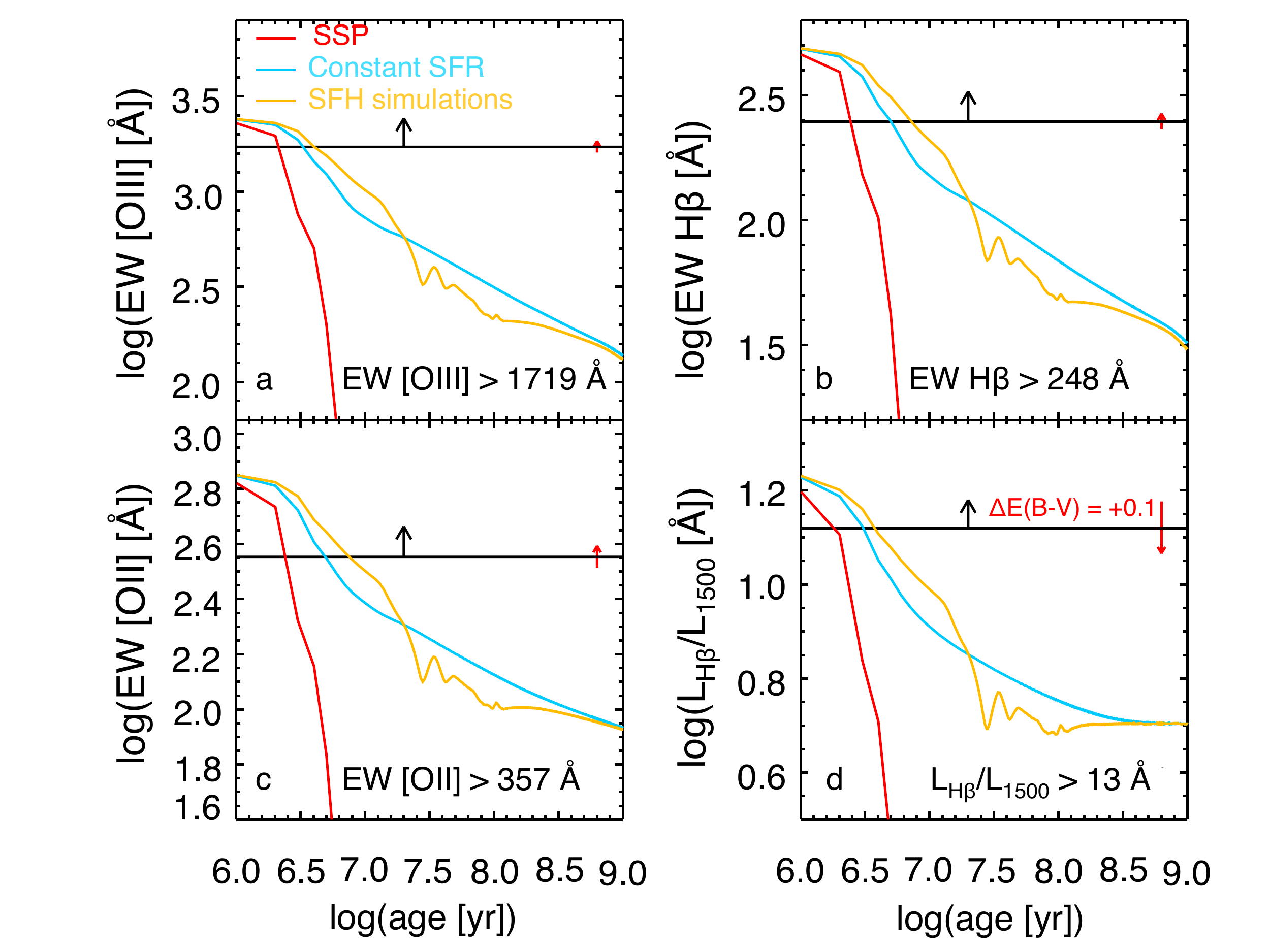}
\caption{Constraints on the clump's age from reddening corrected, rest-frame, emission line EWs.
Lower limits on the clump EW (black solid line) of
[OIII], H$\beta$, [OII] (\textbf{a} - \textbf{c}) and the ratio
between the H$\beta$ luminosity and the 
continuum at 1500 $\mathrm{\AA}$ (\textbf{d}) are compared with theoretical tracks. A Salpeter initial mass function is 
assumed and different star formation histories (SFHs) are compared 
(single burst, constant star formation rate, and a SFH predicted by simulations$^{9}$, Figure 4). The effect of reddening ($\Delta$E(B-V) = +0.1) is indicated in each panel (red arrow). The age of the clump is constrained to be $< 10$ Myr.}
\label{fig:ew}
\end{figure}

\newpage

\begin{figure}[h!]
\centering
\includegraphics[width=\textwidth]{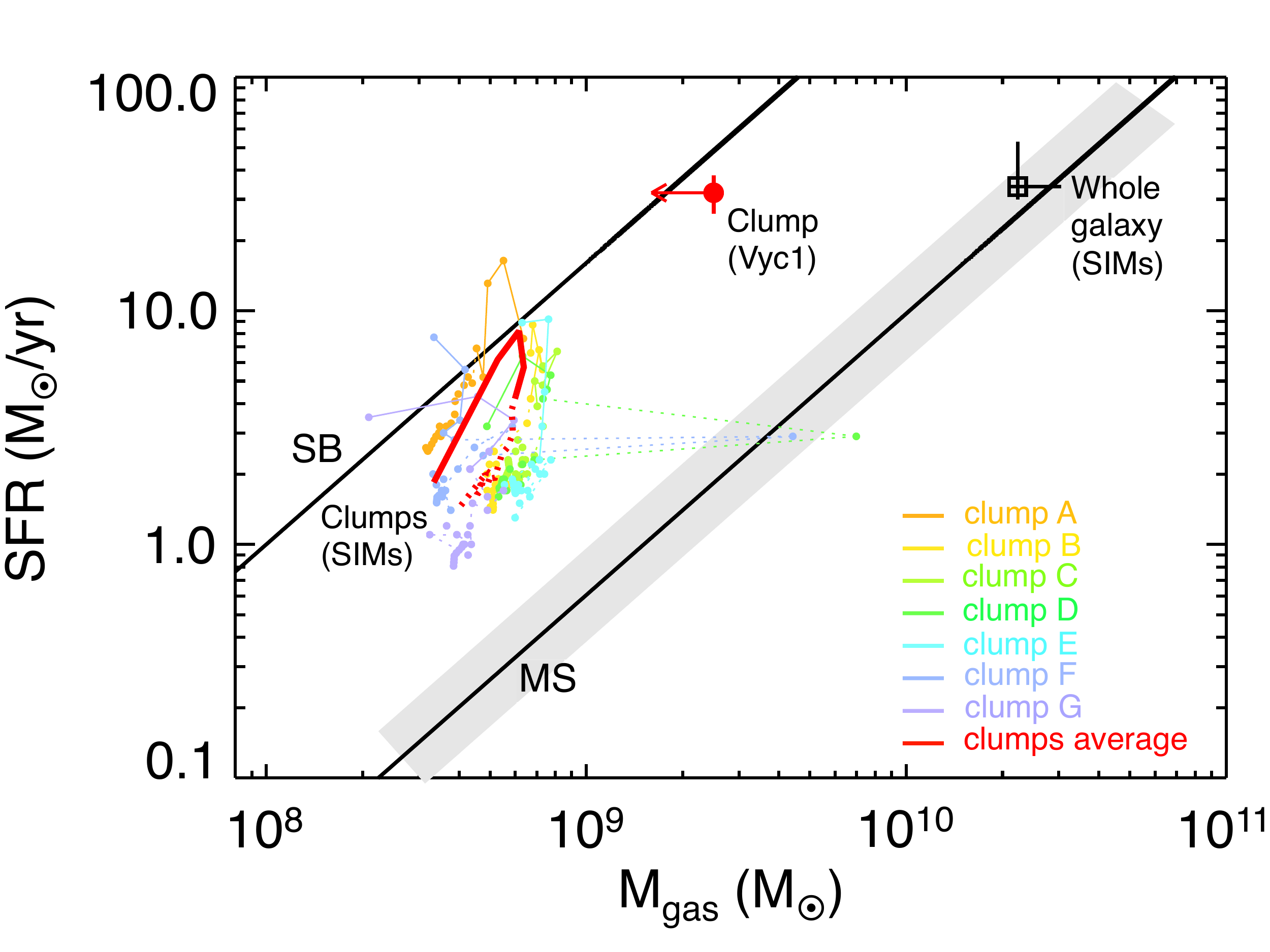}
\caption{Schmidt-Kennicutt plane. We compare the trends for starbursts and MS galaxies$^{29}$
(black solid lines and shaded region indicating the 0.2 dex dispersion of the MS) with the location of observed (red filled circle with s. d. error bars) and simulated clumps (colored dots connected with lines; solid and dashed lines for ages $\leq 30$ Myr and $> 30$ Myr, respectively). The average location of simulated clumps (red, thick line) and host simulated galaxy (square with s. d. error bars) are shown. Sudden variations in clumps' $\mathrm{M_{gas}}$ are likely due to the accretion of gas-rich
clouds or small clumps.}
\label{fig:sk}
\end{figure}

\newpage

\begin{figure}[h!]
\centering
\includegraphics[width=\textwidth]{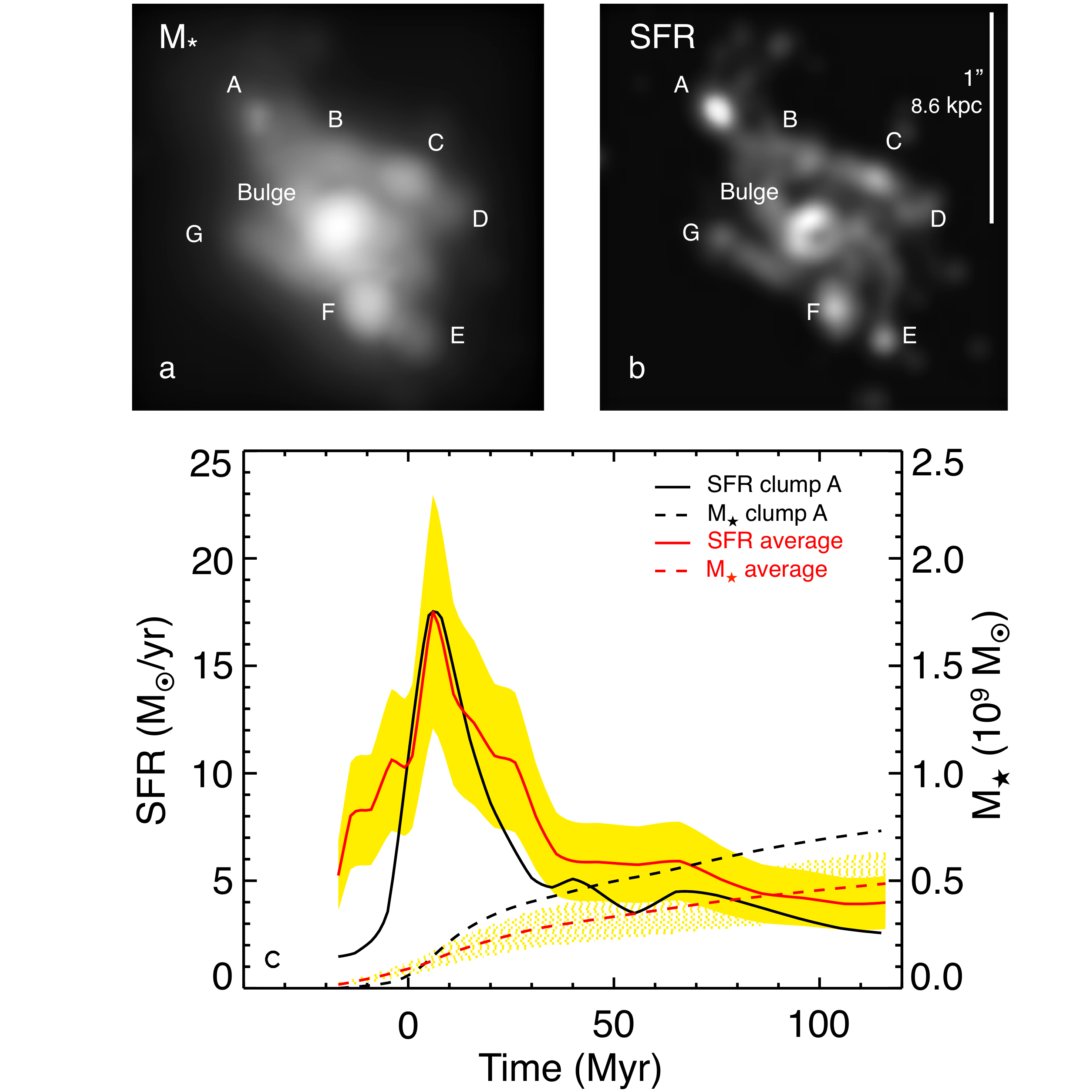}
\caption{Numerical simulations of a high-redshift clumpy galaxy seen face-on. Maps of stellar mass and SFR are shown at \textit{HST}-like resolution (\textbf{a}, \textbf{b}). All clumps have elevated SFR compared with $\mathrm{M_\star}$, but this property is extreme for clump A, observed 12 Myr after its formation. Panel \textbf{c} shows the time evolution of the SFR and $\mathrm{M_\star}$ for clump A and for all the clumps (each star formation history is arbitrarily shifted in time to align the SFR peaks). All clumps experience an internal burst of star formation before evolving into a long-lasting regulated regime within 20 Myr. Yellow shaded regions indicate s.d. uncertainties.}
\label{fig:simulations}
\end{figure}

\newpage

\textbf{METHODS}
\\
\\
\noindent
\textbf{Emission line maps}
\\
The 16 \textit{HST}/WFC3 orbits of G141 slitless spectroscopy,  taken along three position
angles ($\sim0$, -30, +15 degrees)$^{19}$, were reduced
with aXe$^{31}$. Residual
defects (bad pixels, cosmic
ray hits, etc.) were removed with L.A.Cosmic$^{32}$. 
Two dimensional-spectra were background subtracted with
SExtractor$^{33}$, and the continuum emission of the main target and 
surrounding sources (including higher and lower order dispersion
spectra) were removed fitting their aXe continuum models with free normalization (Figure 1).

Astrometrically calibrated emission line
maps were obtained by  cross-correlating the spectral images of [OIII] (the brightest line) with the
three different position angles. 
This is preferred than cross-correlating with the continuum image since our target has different broad-band and line morphologies. 
For  this step, the spectral images were combined with the IRAF task WDRIZZLE$^{34}$, weighting each
single orientation by its exposure time. The astrometry of the \hb ~and [OII] emission maps was tied to that of  [OIII]. The resulting redshift of agrees accurately with Subaru/MOIRCS longslit spectroscopy$^{35}$.

The [OIII] doublet is resolved at the spectral resolution of our data for relatively compact galaxies. We removed the [OIII]$\lambda$4959\AA
~component modelling the combined emission line images with GALFIT$^{36}$ using an effective point-spread function (PSF) consisting of a main lobe for the 5007\AA ~line and three fainter ones.
\\
\\
\noindent
\textbf{Clump continuum emission}
\\
Visual inspection of  the  multi-band \textit{HST} imaging did not reveal any evidence of the clump, 
and  the evaluation   of their isophotal contours did not show disturbances at its location. Thus we searched  for its presence modelling the imaging 
with GALFIT  (Extended Data [ED]~Figure 3).  A single S\'ersic$^{37}$ profile provided a simplified fit, leaving strong positive and negative residuals near the expected position of the clump. Such a pattern is a systematic effect due to the presence of clumps at the outskirts of the galaxy major axis, as they are not symmetrically located with respect to the nucleus, resulting in an effective bending of the galaxy isophotes.  Masking the external regions and fitting the
central part of the galaxy with a single S\'ersic profile left negligible residuals ($\lesssim$~5\%). As a further check, we fitted the direct images with the
Multi-Gaussian Expansion parametrization (MGE) algorithm$^{38}$,  fitting average azymuthal light
profiles with ellipsoidal isophotes to the central part of the galaxy. The residuals are negligible ($\lesssim$~5\%). 
Analogous residuals resulted using three spatially offset  S\'ersic profiles: one centered at the barycenter
of the stellar light (as determined by SExtractor from the F140W image) and other two,
an order of magnitude fainter,  to the top left and bottom right. This is our best fit (baseline) model for the galaxy continuum. 

This  three-component fit is a technical solution adopted due to the irregular morphology of our target, typical of clumpy high-z disks, and should not mislead to concluding 
that the galaxy is an ongoing merger. In this regard, 
we classified the galaxy as a disk based on the  Asymmetry and M$_{20}$ parameters measured on stellar mass maps derived from pixel-to-pixel spectral energy distribution (SED) fitting$^{39,40, 41}$ (ED~Figure 4), a diagnostics calibrated with MIRAGE numerical simulations$^{42}$. Finally, the F105W/F140W ratio (ED Figure 2) provides no evidence for a bulge.

Limits on the clump continuum  were obtained with simulations,  injecting PSF components
 at approximately the same isophotal level of the expected clump position,
and fitting them together with our baseline model.
From these estimates we subtracted the contribution of emission 
lines ([OIII] and \hb\ for F140W, [OII] for F105W), obtaining a factor of 2 (1.3) deeper flux upper limits for F140W (F105W). 
Normalizing a series of Starburst99 stellar population synthesis models$^{43}$ with different stellar ages to the most constraining (F105W) upper limit
allowed us to refine the F140W limit, which is relevant for calculating the  \oiii~and \hb~emission line EWs (ED~Figure 5).
\\
\\
\noindent
\textbf{Clump offset from the galaxy nucleus}
\\
The clump is offset from the galaxy center: the observed distance between the point-like  [OIII] emission  and the barycenter of the galaxy is 1.6~kpc,
with formally negligible measurement error. However, systematic uncertainties exist, related to the astrometric calibration of  the direct imaging and slitless data, and to the stability of the wavelength solution. 
We estimated the systematic uncertainties along the dispersion direction evaluating the  distribution of  differences  between the measured and expected wavelengths of bright emission lines (\ha ~and \oiii) of the full survey data.  Comparing the position of galaxies in the direct imaging with that of the continuum
emission in the grism data we evaluated the systematics in the cross-dispersion direction.
For each orientation of the grism we imposed:
\begin{equation}
\mathrm{\chi_{red}^2 = \frac{1}{N_{dof}} \sum_{i=1}^N
  \left(\frac{s_{meas,i}-s_{exp}}{\sqrt{\sigma_{P,i}^2 + \sigma_A^2}}\right)^2 = 1}
\label{eq:chisquare}
\end{equation}
where $\mathrm{N_{dof}}$ are the  degrees of freedom, $\mathrm{s_{meas,i}}$ and $\mathrm{s_{exp}}$ are respectively the measured
and expected positions of the emission lines (or the continuum) of each galaxy and $\sigma_\mathrm{P}$ and $\sigma_{\mathrm{A}}$ indicate respectively the formal measurement errors on the emission lines (or continuum) positions and the astrometric uncertainties.
Average systematic
uncertainties are $\sigma_A=0.067''$ ($\sigma_A=0.035''$) along the dispersion (cross-dispersion) direction. We computed the uncertainties along the
right ascension and declination directions projecting along the orientation of each dataset. Since the final, astrometrically calibrated emission
line maps are the weighted average of
three different orientations, we estimated  the total uncertainties assuming that the errors ($\epsilon_i$) in each orientation are
independent:
\begin{equation}
\mathrm{\epsilon = \frac{\sqrt{\Sigma_{i=1}^3 (t_i\, \epsilon_i)^2}}{\Sigma_{i=1}^3t_i}}
\end{equation}
where $\mathrm{t_i}$ are the exposure times. 
The clump offset is detected at 7.6$\sigma$ and its projected distance from the galaxy nucleus (defined as the barycenter of the stellar light) is $1.6\, \pm \,0.3$ kpc. If we had chosen the light peak of the direct images as the nucleus, the offset would be comparable in magnitude and significance. We prefer the light barycenter definition as it coincides with the peak of the mass map (ED Figure 2).

To determine the deprojected distance, the axial ratio of the galaxy and the angle $\mathrm{\theta}$ between the galaxy major axis and the clump-nucleus direction are needed. We estimated them from the range of solutions obtained modelling the direct images and the mass map with GALFIT and considering the outer isophotes of the PSF-deconvolved galaxy. To further account for systematic effects we also considered plausible uncertainties in the PSF derivation, and further estimates based on the MGE software as an alternative to GALFIT. Given an axial ratio 0.21 $\leq$ q $\leq$ 0.35 (inclination i $\sim$ 70 -- 78$^\circ$) and 48 $\leq \mathrm{\theta} \leq$ 52$^\circ$, we computed a “maximum plausible range” for the deprojected distance of the clump from the nucleus of $3.6 \leq$ d $\leq  6.2$ kpc, beyond the galaxy effective radius R$_{\mathrm{e}}$ = 2.8 $\pm$ 0.4 kpc (ED Table 1). We did not accounted for the disk thickness: this uncertain correction could imply larger deprojected distance by 10 -- 15\% (for a typical thickness of a few hundreds of pc). 
\\
\\
\noindent
\textbf{Dust reddening} 
\\
Estimating emission line luminosities and SFRs requires dust extinction
corrections (this is less relevant, though, for emission line EWs, affected only by the differential line versus continuum reddening). 
We used stellar population modelling of the UV-to-NIR
galaxy SED$^{44}$, assuming the Calzetti et al.$^{45}$ reddening law and constant SFHs to measure the stellar continuum reddening. We converted this measure into nebular reddening using 
 $\mathrm{E(B -  V)_{nebular} = E(B - V)_{continuum}/0.83}^{46}$,  obtaining  $\mathrm{E(B -
  V)_{nebular}}= 0.30^{+0.09}_{-0.07}$. 
Independent  estimates of the nebular reddening were also
obtained based on emission line ratios: (i) \ha/\hb, assuming case B
recombination conditions$^{47}$; (ii) \oii/\ha, assuming an intrinsic ratio of
1$\, ^{48}$, and (iii) \oii/\hb, with intrinsic ratio estimated following the previous points. For these estimates we
used \ha\ fluxes from MOIRCS, \oii\ from WFC3 and \hb\ from the weighted average of MOIRCS and WFC3, obtaining:
 $\mathrm{E(B - V)_{H\alpha/H\beta} = 0.24\pm0.12}$;
$\mathrm{E(B - V)_{[OII]/H\alpha} = 0.32\pm0.11}$; \\ $\mathrm{E(B -
  V)_{[OII]/H\beta} = 0.40\pm0.25}$. The average of these estimates is nearly identical to that from the stellar continuum. We therefore adopt  $\mathrm{E(B -  V)_{nebular}}= 0.30$.
  
For the clump a reddening estimate can be obtained using  WFC3, from the ratio of   the \oii ~and \hb
~line fluxes. We derived  a fairly noisy measurement that is 
consistent with that of the whole galaxy ($\mathrm{E(B-V)_{[OII]/H\beta,clump} = 0.24 \pm 0.37}$). While formally
this is also consistent with zero attenuation towards the clump,  this is unlikely as the galaxy is highly inclined. To improve the estimate of the reddening affecting the clump, we attempted a derivation of the \ha\ flux
of the clump in the MOIRCS data, decomposing the 2D spectrum with a PSF-like component  for the clump and a single
S\'ersic profile accounting for the host galaxy disk, finding \ha$=7\pm2\times10^{-17}$~erg~s$^{-1}$~cm$^{-2}$, $\sim50$\% of the
galaxy \ha\ emission$^{35}$. Averaging the reddening  estimates from  \ha/\hb,  \ha/\oii\ and \oii/\hb, we obtained $\mathrm{E(B -  V)_{nebular,clump}}=0.55\pm0.20$, 
consistent with the reddening of the host galaxy. We thus assumed that the clump nebular reddening is identical to that of the parent galaxy, consistently
with the literature$^{49}$. ED~Figure 2 shows the observed F606W/F105W ratio, probing the stellar continuum reddening, which is homogeneous over the galaxy.  The optical attenuation ($\mathrm{A_V}$) at the clump position is similar to that at the galaxy nucleus within 0.1 -- 0.2~mag,  and close to the galaxy average. The position of the galaxy nucleus (measured as the light barycenter, light peak, or with GALFIT) is stable and not changing with the wavelength from F606W to F105W and F140W. \oiii ~and F105W continuum should be affected by a similar attenuation and much less than the F606W continuum. Together with the 
flatness of the reddening map, this demonstrates that the clump emission lines are not an artifact due to reddening modifying the galaxy nucleus position, as an even stronger effect would be seen in F606W. Reddening correcting the emission line maps and the imaging does not significantly alter the nucleus-clump distance. Adopting  the Cardelli et al.$^{50}$ extinction law  (see, e.g., Steidel
et al.$^{51}$)  would produce reddening values $\lesssim$ 15\% higher,  consistent within the uncertainties.
\\
\\
\noindent
\textbf{Discarding the AGN, shock, transient, and low-metallicity region hypotheses}
\\
The galaxy has  three Chandra photons (1~soft and 2~hard; $\sim 2 \sigma$ detection) in  
 146~ks data  giving  $\mathrm{L_{2-10\,keV} \sim2.9\times 10^{42}}$ \es\ (photon index $\Gamma=1.8$). This is 10 times higher
 than expected from the galaxy star formation$^{52}$. If an active galactic nucleus (AGN) were present, it would produce$^{53}$ an 
 \oiii ~luminosity $\sim20$ times fainter than that of the clump. 
In ED~Figure 6 both the entire galaxy  and the clump  are located in the BPT$^{54}$ diagram (we conservatively use $\mathrm{[NII]_{clump} \lesssim
[NII]_{galaxy}}$). The
emission line ratios  are  consistent with star-forming galaxies at z $\sim$ 2$^{51}$. 
The \oiii/\nii\ $< 2.8$ upper limit is also much lower than typically observed in Type~1 AGNs$^{55}$.
The high EW further disfavors the hypothesis of
an off-nuclear AGN, since  AGNs typically have $\mathrm{EW_{[OIII]}<500\, \AA}^{56}$. 
Besides, no AGN signature was found from the galaxy SED,
  and no excess possibly arising from nuclear accretion is detected in our deep
24 $\mu$m--Spitzer, Herschel and VLA  data. 

The clump emission line luminosity is comparable with that of the whole galaxy, hence it cannot be due to shock from external outflows impacting the gas. 
The host SFR  would generate $\sim30$~times weaker  galaxy-integrated, shock-excited line
luminosities$^{57}$. The brightest shock powered off-nuclear clouds in local IR luminous galaxies are  $>$ 50~times weaker$^{58}$.
Explicit calculations$^{59}$ for $\mathrm{z}=2$ galaxies using appropriate wind mass loads$^{60,61}$ and velocities$^{62}$ lead to analogous conclusions. The  kinetic energy available in winds cannot account for the clump line luminosities.

There is not evidence for substantial line luminosities variability over a $\sim$ 3\,yr timescale. \textit{HST}/WFC3 G141 spectroscopy was obtained in June and July 2010, and 
MOIRCS spectroscopy in April 2013 (ED Table 2). Despite their lower resolution ($0.6''$~seeing),  MOIRCS spectra show the bright, compact \oiii\ and \ha\ emissions from the clump, with a consistent flux. 

Low-mass ($<10^9\, \mathrm{M_\odot}$), very metal poor galaxies ($\mathrm{Z \sim 0.1\, Z_{\odot}}$) can display  extremely high EW emission lines$^{63}$. 
Our target is substantially more massive and metal rich: 
using the \oiii/\oii ~ratio we estimated$^{64}$  $\mathrm{Z
\sim 0.4\pm 0.1 \, Z_{\odot}}$ and  $\mathrm{Z \sim 0.6\pm 0.2 \, Z_{\odot}}$, for the clump and galaxy, respectively (ED~Figure 4, ED Table 1). 
\\
\\
\noindent
\textbf{Constraining the age of the clump}
\\
We computed the time evolution of the
\hb ~EW using stellar population synthesis models$^{43}$, adopting
$\mathrm{Z} = 0.4\, \mathrm{Z_{\odot}}$, a Salpeter$^{65}$ initial
mass function (IMF), and three
different SFHs: an instantaneous burst,  constant
star formation and a SFH obtained from our
hydrodynamic simulations (Figure 3). All models show  high EWs at young ages ($\mathrm{log(EW)
  \gtrsim 2}$, independent of the SFH), which drop quickly for the
 instantaneous burst and more smoothly
in the other cases. We converted the \hb ~EW into the expected \oiii ~and \oii ~EWs (Figure 2), 
assuming the \oiii/\hb ~ratio of $\mathrm{z}=2$ star-forming galaxies$^{51}$  and an \ha/\oii ~luminosity ratio of 1$^{48}$.
Comparing with the EW lower limits we inferred an age $<10$~Myr for the clump (Figure 2). 

The directly measured continuum upper limits (rather than the more stringent ones from synthetic spectra; ED Figure 5), give an age $\lesssim15$~Myr. 
Decreasing the adopted E(B-V) reddening by 0.1~dex would reduce the line EW lower limits by $\sim 0.05$~dex only, but increasing the $\mathrm{L_{H\beta}}$/$\mathrm{L_{1500}}$ limit by $\sim$ 0.1~dex, hence with hardly any effect on the age constraints. 
Changing the metallicity by 1.6~dex produces a 0.2~dex
age difference only. Similarly, the age  remains unchanged adopting
 e.g., a Kroupa$^{66}$ or a Scalo$^{67}$
IMF, and a top-heavy one produces EWs  0.2~dex
higher.
\\
\\
\noindent
\textbf{SFR estimate}
\\ 
 The  SFR of the whole galaxy  was determined from the total \ha
~luminosity  from the MOIRCS spectroscopy, assuming the
standard Kennicutt conversion$^{68}$, resulting in  $\mathrm{77\pm 9 \,
  M_\odot}$~yr$^{-1}$, in agreement with that
from SED fitting ($\mathrm{\sim 85 \, M_\odot}$~yr$^{-1}$ with an uncertainty of 0.2 dex, ED Table~1).

The line luminosity to SFR time dependent conversion at young ages was computed  using Starburst99 adopting the 
SFH  from our numerical simulations  (ED~Figure 7).
At $\mathrm{t}=10$~Myr this is 20\% higher than Kennicutt. Averaging the estimates from $\mathrm{H\beta}$, \oii\ and $\mathrm{H\alpha}$ we obtained 
$\mathrm{SFR}=32\pm6\, \mathrm{M_\odot}$~yr$^{-1}$ for the clump, where the error includes the uncertainties associated with emission line luminosities and reddening. 
\\
\\
\noindent
\textbf{Stellar mass estimate}
\\
Assuming the average mass-to-light ratio (M/L) of the host galaxy (ED~Figure 2), the flux upper limit on
the continuum  emission of the clump implies $\mathrm{M_{\star} \lesssim 3\cdot
10^8\, M_{\odot}}$. Using the M/L ratio from the clump SFH (ED~Figure 7) gives $\mathrm{M_{\star} \lesssim 2.1\cdot
10^8\, M_{\odot}}$. 
Normalizing the simulations to the  observed \hb\  luminosity yields $\mathrm{M_{\star} \sim 3.9 \cdot 10^8\,
  M_{\odot}}$, 
 consistent with the previous estimates
given the uncertainties 
(a factor $\sim2$, mainly due to the gas fraction of simulated galaxies and the details
of feedback and stellar mass loss modelling  at small scales; note that the simulated clumps appear to be slightly less massive than our observed one, on average, but their observed physical properties and time behaviour is self-similar).
\\
\\
\noindent
\textbf{Gas mass estimate}
\\
 We  inferred an upper limit to the gas mass in the clump   from the Jeans mass ($\mathrm{M_J}$) of the galaxy, which is close to 
the maximum gas mass that can collapse in a rotation disk$^{2,9,69,70}$.
Assuming a reasonable upper limit for the typical gas velocity dispersion 
in high-z disk galaxies ($\mathrm{\sigma_V} \lesssim 80$ km/s$^{21,22}$), we obtained   $\mathrm{M_{gas}  \lesssim 
{M_J}=2.5\cdot\,10^9 \,M_{\odot}}$. Using the $\mathrm{M_{gas}}$/\hb\ ratio from simulations  leads to a gas mass of  $\mathrm{M_{gas} \sim 2.7 \cdot 10^9\,M_{\odot}}$.
Comparing with older clumps  from the literature$^{13}$, using  our numerical  simulations to relate the physical properties at the ``peak'' and later phases, yields:
\begin{equation}
\mathrm{M_{gas, clump} = \frac{SFR_{clump}}{SFR_{lit}}\cdot M_{\star,lit}\cdot
\left(\frac{M_{gas,young}}{M_{\star,old}}\right)_{sim}\cdot \left(\frac{SFR_{young}}{SFR_{old}}\right)_{sim}}
\label{eq:gas_mass}
\end{equation}
where $\mathrm{SFR_{lit}}$ and $\mathrm{M_{\star,lit}}$ refer to older clumps reported in the
literature. $\mathrm{M_{\star,old,sim}}$ and
$\mathrm{SFR_{old,sim}}$ are for old clumps in the simulations, while $\mathrm{M_{gas,young, sim}}$ and
$\mathrm{SFR_{young, sim}}$ are computed at $\mathrm{t} = 10$~Myr, as our young
clump. This approach leads to $\mathrm{M_{gas} = 3 \cdot 10^9 \,
  M_{\odot}}\, \pm 0.2$ dex,  consistent with the independent estimates discussed above.  This agreement supports  SFHs with an initial 
  burst as predicted by our simulations.
  
The  Schmidt-Kennicutt relation can be used to provide alternative estimates, based on the clump SFR.
The relation for MS galaxies would imply that $\sim50$\% of the total gas in the
galaxy is collapsing in an ultra-compact region, which appears hardly believable,  confirming that this young clump has higher SFE.  Assuming instead the starburst-like relation, we obtained
$\mathrm{M_{gas} = 2\cdot 10^{9\, +0.36dex}_{\, -0.23dex} \, M_{\odot}}$, consistent with
the previous estimates. 

Considering an even younger age for the clump, as permitted by the upper limit $\mathrm{t < 10\, Myr}$, would return higher absolute
and specific SFRs (ED~Figure 7; i.e. higher SFR/L$_{\mathrm{H\beta}}$) confirming the starburst behaviour of the clump during the formation
phase.
\\
\\
\noindent
\textbf{Dynamical time estimate}
\\
Measuring the FWHM of
the \ha  ~line detected in the MOIRCS longslit spectroscopic data, we
determined a first upper limit on the gas velocity of the clump
$\mathrm{v_{FWHM} \lesssim 450\, km\, s^{-1}}$ (the MOIRCS instrumental resolution). Given the upper limits on the radius of the clump ($\mathrm{R < 500\,
pc}$) and on its dynamical mass ($\mathrm{M_{dyn} =
M_{gas} + M_\star \lesssim 2.8\cdot 10^9\, M_\odot}$), we then refined our estimate to $\mathrm{v_{FWHM} < \sqrt{M_{dyn}\, G/R}
\sim 200\, km\, s^{-1}}$ (G is the gravitational constant), consistent with clump
velocities typically observed in high-redshift galaxies$^{17}$. This leads to a dynamical timescale  $\mathrm{t_{dyn} =
2\pi R/(v_{FWHM}/2) \sim 29\, Myr}$, reasonably in agreement with
the free-fall time of the clump $\mathrm{t_{ff} \sim \sqrt{R^3/M_{dyn}}
  \sim 17\, Myr}$.
\\
\\
\noindent
\textbf{Clump formation rate and lifetime}
\\
The visibility window of the young phase
can be defined as the time during which the EW is above a given threshold, as predicted by stellar population synthesis models. 
For our clump this ranges between $\sim$ 5 Myr  (instantaneous
burst) and $\sim$ 10 Myr for SFH from simulations. We used an average visibility window of 7 Myr. 

Knowing the visibility window and the observed number of ``formation events'' per galaxy, the ``clump formation rate'' can be estimated and comparing it to the average number of descendants observable per galaxy (virtually all old) yields the average lifetime of the clumps. 

We considered all galaxies in our survey with: (i) $\mathrm{M_{\star} > 8.5 \cdot 10^9\,
  M_{\odot}}$ (mass completeness, coinciding with the minimum
mass that a galaxy should have to host such a massive clump assuming
a gas fraction of $\sim$ 50\%); (ii) $\mathrm{M_{\star} < 2\cdot 10^{11}\, M_{\odot}}$  (\oiii ~emission becomes too weak at higher masses$^{71}$); and (iii)  a redshift $\mathrm{1.2 < z < 2.4}$ (\oiii\ emission lying inside the wavelength range of the grism).
57 galaxies are selected in this way. With one ``forming clump'' detected, this corresponds to a ``clump formation rate'' of 2.5~Gyr$^{-1}$ per galaxy. 

We considered that in our survey we would have detected all formation events of clumps with $\mathrm{M_{gas} \gtrsim 2.5\cdot 10^9\, M_{\odot}}$.
Considering that almost all the initial gas mass of a given clump is consumed at initial stages to form
stars, the clump stellar mass at late stages can be approximated to the gas mass at initial collapse, independently of the age of the clump, as supported
by our numerical simulations. Typically there are $\sim$ 1 -- 2 clumps per galaxy above such mass
threshold$^{13,17,18}$, giving an average lifetime of 500~Myr. 

To compute the (large) associated
uncertainty,  we considered (i) the Poisson error associated to our single object discovery, (ii) the Poisson error for older clumps from the literature; and (iii)
the visibility window uncertainty. The asymmetric $1\sigma$
uncertainties that we inferred  are +0.74 dex and -0.55 dex. The lower-envelope of the 1$\sigma$ range of the lifetime estimate is not 
far from the upper range of lifetimes suggested by models in which clumps suffer from strong feedback (50 -- 100~Myr). 
However, our estimate is likely a lower-limit.
The derived lifetime could be
affected by a ``discovery bias'', since other high-redshift  spectroscopic surveys  (e.g., SINS, 3D-HST)
have not yet  reported  the observation of a similar giant  young clump. Furthermore, there are indications$^{35}$ that our target galaxy is living 
in a gas-enriched environment which could also have anomalously increased the SFR and thus the ``clump formation rate''. 
This suggests that the observation of a newly giant clump could be an even rarer event
than what appears from our data, and that the true average clump lifetime could be
longer than estimated here.
\\
\\
\noindent
\textbf{Code availability}
\\
The RAMSES code used to generate our simulations is available at:
\\
 http://www.ics.uzh.ch/$\sim$teyssier/ramses

\clearpage 

\textbf{Additional references}
\\
31. K\"ummel, M. et al. The slitless spectroscopy data extraction
software. \textit{PASP} \textbf{121} 59-72 (2009)
\\
32. van Dokkum, P. et al. Cosmic-ray rejection by Laplacian edge
detection. \textit{PASP} \textbf{113} 1420-1427 (2001)
\\
33. Bertin, E. \& Arnouts, S. SExtractor: software of source extraction. \textit{Astron. \& Astrophys. (Suppl.)} \textbf{117} 393-404 (1996)
\\
34. Fruchter, A. S. \& Hook, R. N. Drizzle: a method for the linear
reconstruction of undersampled images \textit{PASP} \textbf{114} 144 (2002)
\\
35. Valentino, F. et al. Metal deficiency in cluster star-forming
galaxies at z = 2, \textit{Astrophys. J.}, in press, arXiv:1410.1437
\\
36. Peng, C. Y. et al. Detailed decomposition of galaxy images. II. Beyond
axisymmetric models. \textit{Astronom. J.} \textbf{139} 2097-2129 (2010)
\\
37. S\'ersic, J. L. et al. Influence of the atmospheric and
instrumental dispersion on the brightness distribution in a
galaxy. \textit{Boletin de la asociation astronomica de astronomia}
\textbf{6} 41 (1963)
\\
38. Cappellari, M. Efficient multi-Gaussian expansion of galaxies
\textit{Mont. Notic. R. Astron. Soc.} \textbf{333} 400 - 410
\\
39. Conselice, C. J. The relationship between stellar light
distributions of galaxies and their formation
histories. \textit{Astrophys. J. (Supp.)} \textbf{147} 1-28 (2003)
\\
40. Lotz, J. M. et al. A new non-parametric approach to galaxy
morphological classification. \textit{Astronom. J.} \textbf{128}
163-182 (2004)
\\
41. Cibinel, A. et al. A physical approach to the identification of high-z mergers: morphological classification in the stellar mass domain. \textit{Astroph. J.}, submitted
\\
42. Perret, V. et al. Evolution of the mass, size, and star formation
rate in high redshift merging galaxies. \textit{Astron. \& Astrophys.}
\textbf{562} 1 (2014)
\\
43. Leitherer, C. et al. Starburst99: synthesis models for galaxies
with active star formation. \textit{Astrophys. J. (Suppl.)}
\textbf{123} 3-40 (1999)
\\
44. Strazzullo, V. et al. Galaxy evolution in overdense environments
at high redshift: passive early-type galaxes in a cluster at z $\sim$
2. \textit{Astrophys. J.} \textbf{772} 118-135 (2013)
\\
45. Calzetti, D. et al. The dust content and opacity of actively
star-forming galaxies. \textit{Astrophys. J.} \textbf{533} 682-695 (2000)
\\
46. Kashino, D. et al. The FMOS-COSMOS survey of star-forming
galaxies at z $\sim$ 1.6. I. \ha ~based star formation rates and dust
extinction. \textit{Astrophys. J. Lett} \textbf{777} 8-14 (2013)
\\
47. Osterbrok, D. E. et al. \textit{Astrophysics of Gaseous Nebulae
  and Active Galactic Nuclei} University Science Books. Mill Valley, CA
\\
48. Kewley, L. et al. \oii ~as a star formation rate
indicator. \textit{Astron. J.} \textbf{127} 2002-2030 (2004)
\\
49. Elmegreen, D. M. et al. Resolved galaxies in the Hubble Ultra Deep
Field: star formation in disks at high redshift \textit{Astroph. J.}
\textbf{658} 763-777 (2007)
\\
50. Cardelli J. A., Clayton, G. C. and Mathis, J. S. The relationship
between infrared, optical, and ultraviolet extinction
\textit{Astrophys. J.} \textbf{345} 245-256 (1989)
\\
51. Steidel, C. C. et al. Strong nebular line ratios in the spectra of
z $\sim$ 2 - 3 star-forming galaxies: first results from
KBSS-MOSFIRE. submitted
\\
52. Persic, M. et al. 2-10 keV luminosity of high-mass binaries as a
gauge of ongoing star formation rate. \textit{Astron. \& Astrophys.}
\textbf{419} 849-862 (2004)
\\
53. Panessa, F. et al. On the X-ray, optical emission line and black
hole mass properties of local Seyfert galaxies. \textit{Astron. \&
Astrophys.} \textbf{455} 173-185 (2006)
\\
54. Baldwin, J. A., Phillips, M. M.,  and Terlevich, R. Classification
parameters for the emission-line spectra of extragalactic objects. \textit{PASP}
\textbf{93} 5-19 (1981)
\\
55. Stern, J. et al. Type 1 AGNs at low z - III. The optical narrow
line ratios. \textit{Mont. Not. R. Astron. Soc.} \textbf{431} 836-857 (2013)
\\
56. Caccianiga, A. et al. The relationship between
[OIII]$\lambda$5007$\mathrm{\AA}$ equivalent width and obscuration in
active galactic nuclei. \textit{Mon. Not. R. Astron. Soc.}
\textbf{415} 1928-1934 (2011)
\\
57. Hong, S. et al. Constraining stellar feedback: shock-ionized gas in nearby starburst galaxies. \textit{Astrophys. J.} \textbf{777} (2013)
\\
58. Soto, K.~T. \& Martin, C.~L. Gas excitation in ULIRGs: maps of diagnostic emission-line ratios in space and velocity. \textit{Astrophys. J. Supp.} \textbf{203} (2012)
\\
59. Binette, L., Dopita, M.~A., and Tuohy, I.~R. Radiative shock-wave theory. II - High-velocity shocks and thermal instabilities. \textit{Astroph. J.} \textbf{297} 476 - 491 (1985)
\\
60. Genzel, R. et al. Evidence for wide-spread active galactic nucleus-driven outflows in the most massive z $\sim$ 1-2 star-forming galaxies. \textit{Astroph. J.} \textbf{796} (2014)
\\
61. Renaud et al. A sub-parsec resolution simulation of the Milky Way: global structure of the interstellar medium and properties of molecular clouds. \textit{Mon. Notic. R. Astron. Soc.} \textbf{436} 1836-1851 (2013)
\\
62. F\"orster Schreiber, N. et al. The Sins/zC-Sinf survey of z $\sim$ 2 galaxy kinematics: evidence for powerful active galactic nucleus-driven nuclear outflows in massive star-forming galaxies. \textit{Astroph. J.} \textbf{787} (2014)
\\
63. Amor\'in et al. Discovering extremely compact and metal-poor,
star-forming dwarf galaxies out to z $\sim$ 0.9 inthe VIMOS Ultra-Deep
survey. \textit{Astron. \& Astrophys.} \textbf{568} 8 (2014)
\\
64. Maiolino, R. et al. AMAZE. I. The evolution of the
mass-metallicity relation at z $> 3$. \textit{Astron. \& Astrophys.} \textbf{488} 463-479 (2008)
\\
65. Salpeter, E. E. The luminosity function and stellar evolution. \textit{Astrophys. J.} \textbf{121} 161-167 (1955)
\\
66. Kroupa, P. \& Boily, C. M. On the mass function of star
clusters. \textit{Mon. Notic. R. Astron. Soc.} \textbf{336} 1188 - 1194 (2002)
\\
67. Scalo, J. M. Two power-law IMF. \textit{Fundam. Cosmic Physics} \textbf{11} 1 (1986)
\\
68.  Kennicutt, R. C. The global Schmidt law in star-forming
galaxies. \textit{Astrophys. J.} \textbf{498} 541-552 (1998)
\\
69. Bournaud, F., Elmegreen, B. Unstable disks at high redshift:
vidence for smooth accreation in galaxy
formation. \textit{Astrophys. J. Lett} \textbf{694} 158-161 (2009)
\\
70. Elmegreen, B., Burkert, A. Accretion-driven turbolence and the
transition to global instability  young galaxy
disks. \textit{Astrophys. J.} \textbf{712} 294-302 (2010) 
\\
71. Zahid, H. J. et al. The universal relation of galactic chemical
evolution: the origin of the mass-metallicity
relation. \textit{Astroph. J.} \textbf{791} 130 (2014)

\clearpage

\begin{figure}[h!]
\centering
\includegraphics[width=\textwidth]{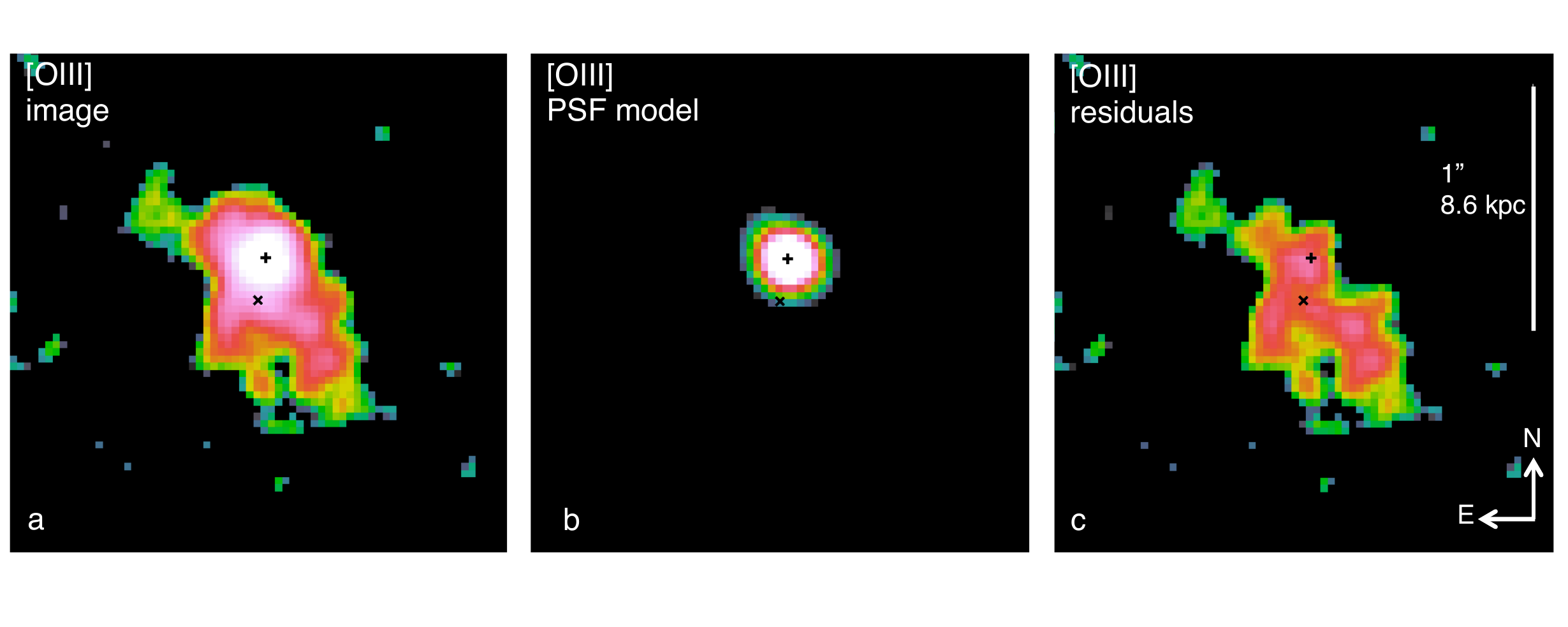}
\caption*{Extended Data Figure 1: GALFIT decomposition of the clump. The [OIII] map (\textbf{a}) and the model of the point source component for the clump (\textbf{b}) are shown. No strong residuals or artifacts are left after removal of the  point source component (\textbf{c}).  The position of the nucleus and of the clump are shown as crosses.
}
\label{fig:OIII}
\end{figure}

\clearpage
\begin{figure}[h!]
\centering
\includegraphics[width=\textwidth]{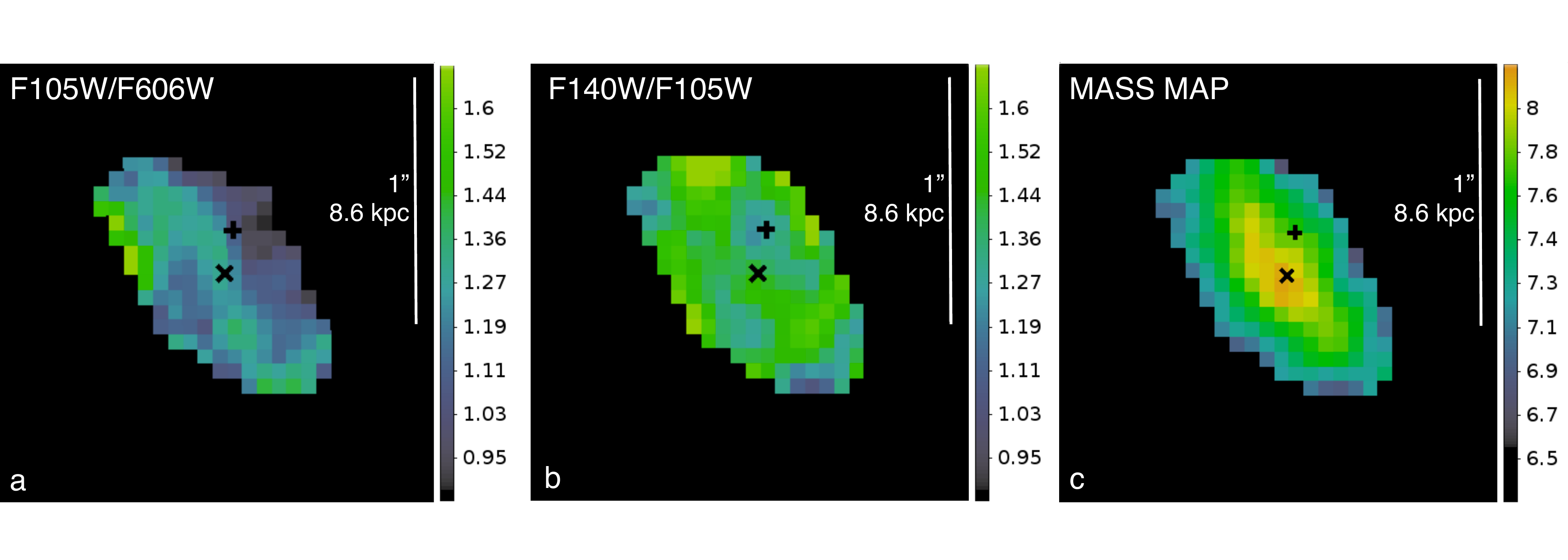}
\caption*{Extended Data Figure 2: Images ratios and mass map. The ratio of F105W/F606W imaging in $\mathrm{F_{\nu}}$ scale (\textbf{a}), a proxy for the dust reddening of the stellar continuum, and F140W/F105W imaging, sensitive to the M/L ratio (\textbf{b}), are shown. The position of the nucleus and the clump are shown as crosses. The maps show only small variations: the observed $\mathrm{F_{\nu}}$ ratios for the nucleus (clump) positions are 1.34 (1.16) for F105W/F606W and 1.39 (1.25) for F140W/F105W. Galaxy-wide ratios are 1.27 and 1.37 for F105W/F606W and F140W/F105W, respectively.  The mass map (\textbf{c}) is shown in units of log$_{10}$(M$_\odot$/pixel).
}
\label{fig:OIII}
\end{figure}

\clearpage

\begin{figure}[h!]
\centering
\includegraphics[scale=0.12]{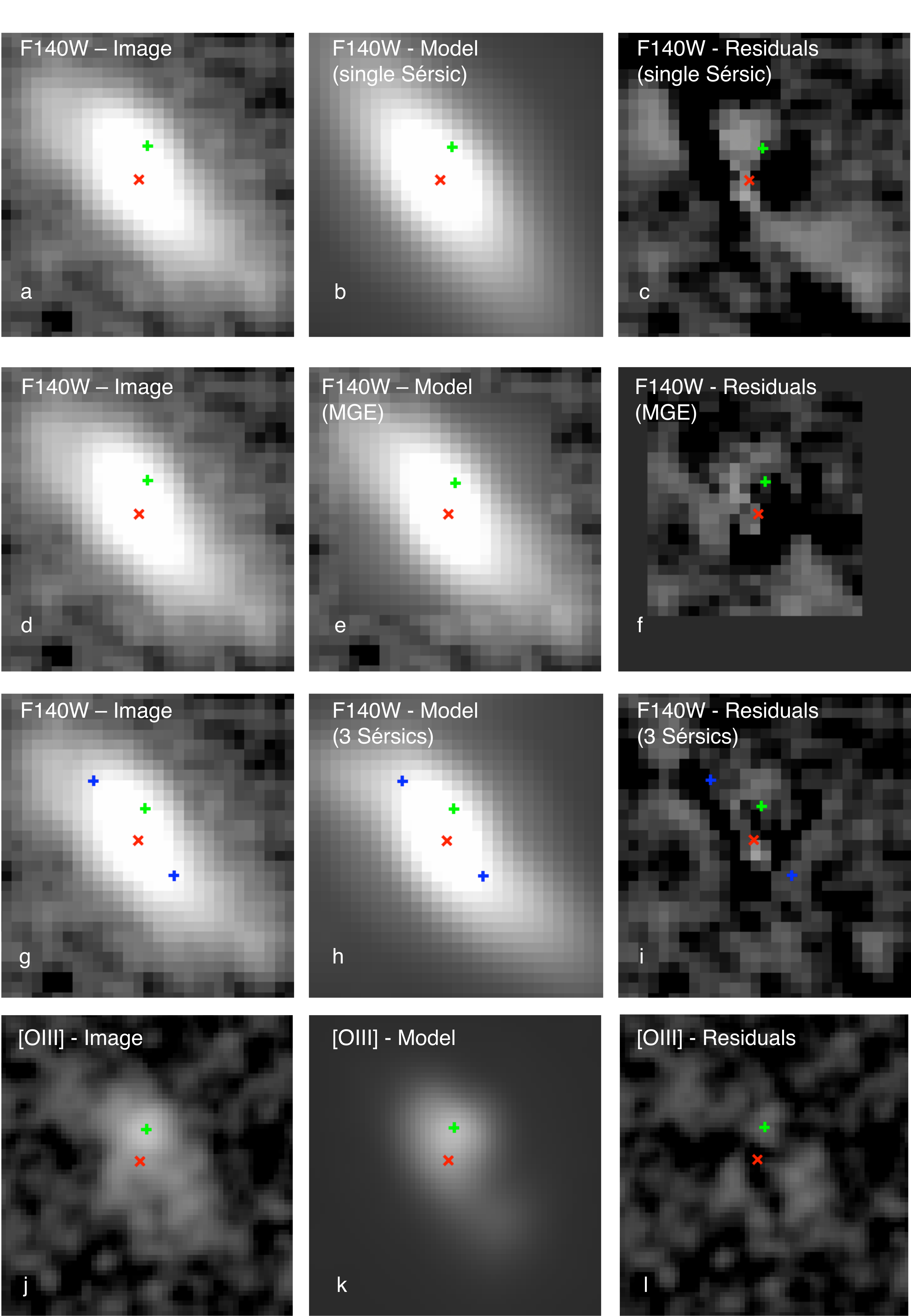}
\caption*{Extended Data Figure 3: Modelling of the galaxy light profile. The F140W direct image and
 \oiii ~emission line map (\textbf{a}, \textbf{d}, \textbf{g}
 and \textbf{j}), the GALFIT models (\textbf{b}, \textbf{e}, \textbf{h}, \textbf{k}) and the residuals
 (\textbf{c}, \textbf{f}, \textbf{i}, \textbf{l}) are
shown. The first row shows the single
S\'ersic profile solution, the MGE model is in the
second row, and our baseline model (the sum of three S\'ersic
profiles; blue crosses mark the additional components) in the third row. The red cross indicates the barycenter of the stellar light and the green one marks the center of the \oiii~
off-nuclear component.}
\label{fig:galfit}
\end{figure}

\clearpage

\begin{figure}[h!]
\centering
\includegraphics[width=\textwidth]{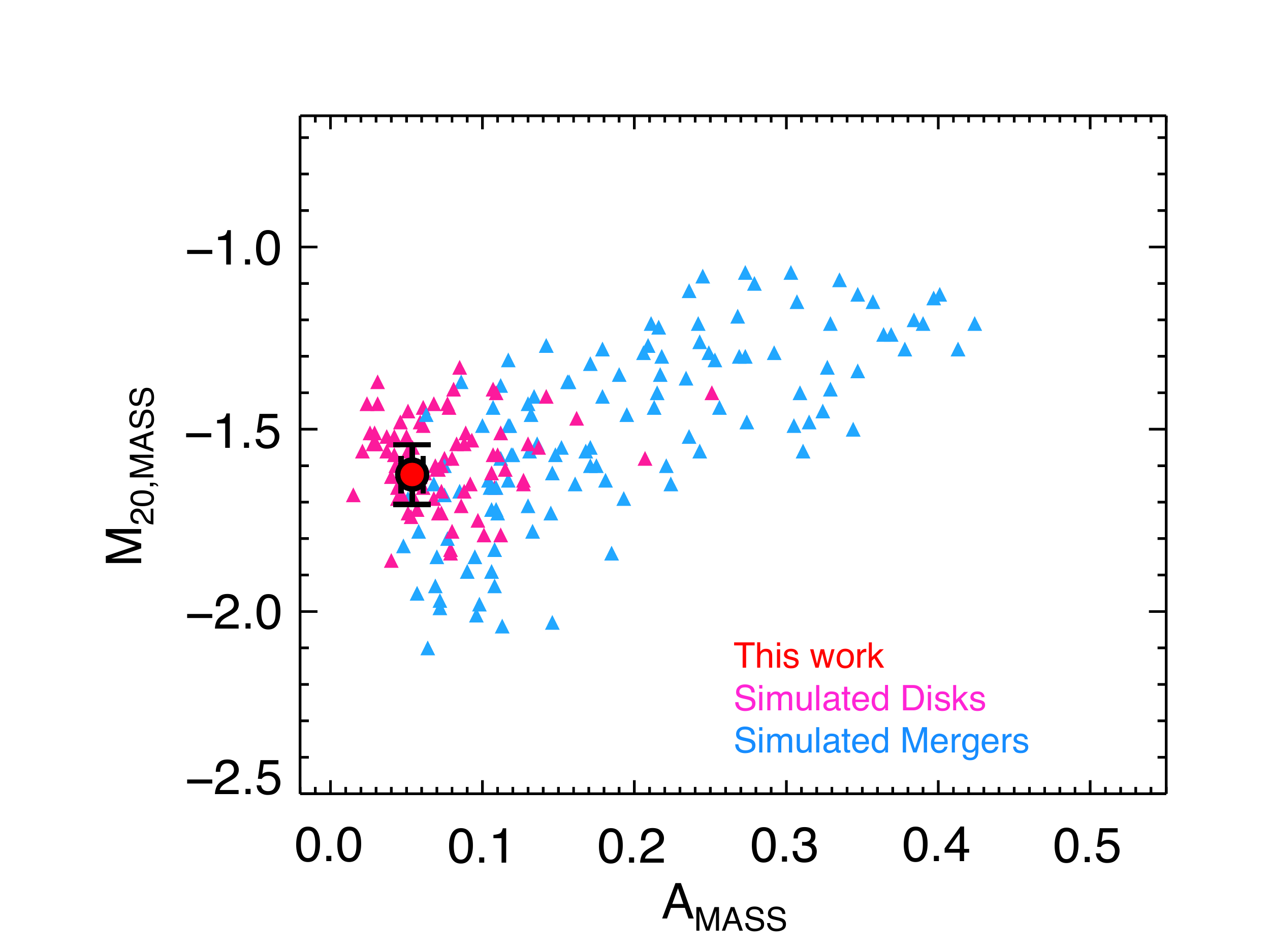}
\caption*{Extended Data Figure 4: The Asymmetry and M$_{20}$ morphological parameters as
  determined from the spatial distribution of the galaxy stellar mass. Pink and light blue triangles
  represent disks and mergers from MIRAGE numerical
  simulations$^{42}$, respectively. The galaxy presented in this work (red filled circle with s. d. error bars) is located in the typical region occupied by disk galaxies$^{41}$ while the vast majority of mergers have higher Asymmetry and/or M$_{20}$ parameters. We note that the figure shows the same number of mergers and disks even if in optical samples mergers are expected to be a minority.}
\label{fig:asymm20}
\end{figure}

\clearpage

\begin{figure}[h!]
\centering
\includegraphics[width=\textwidth]{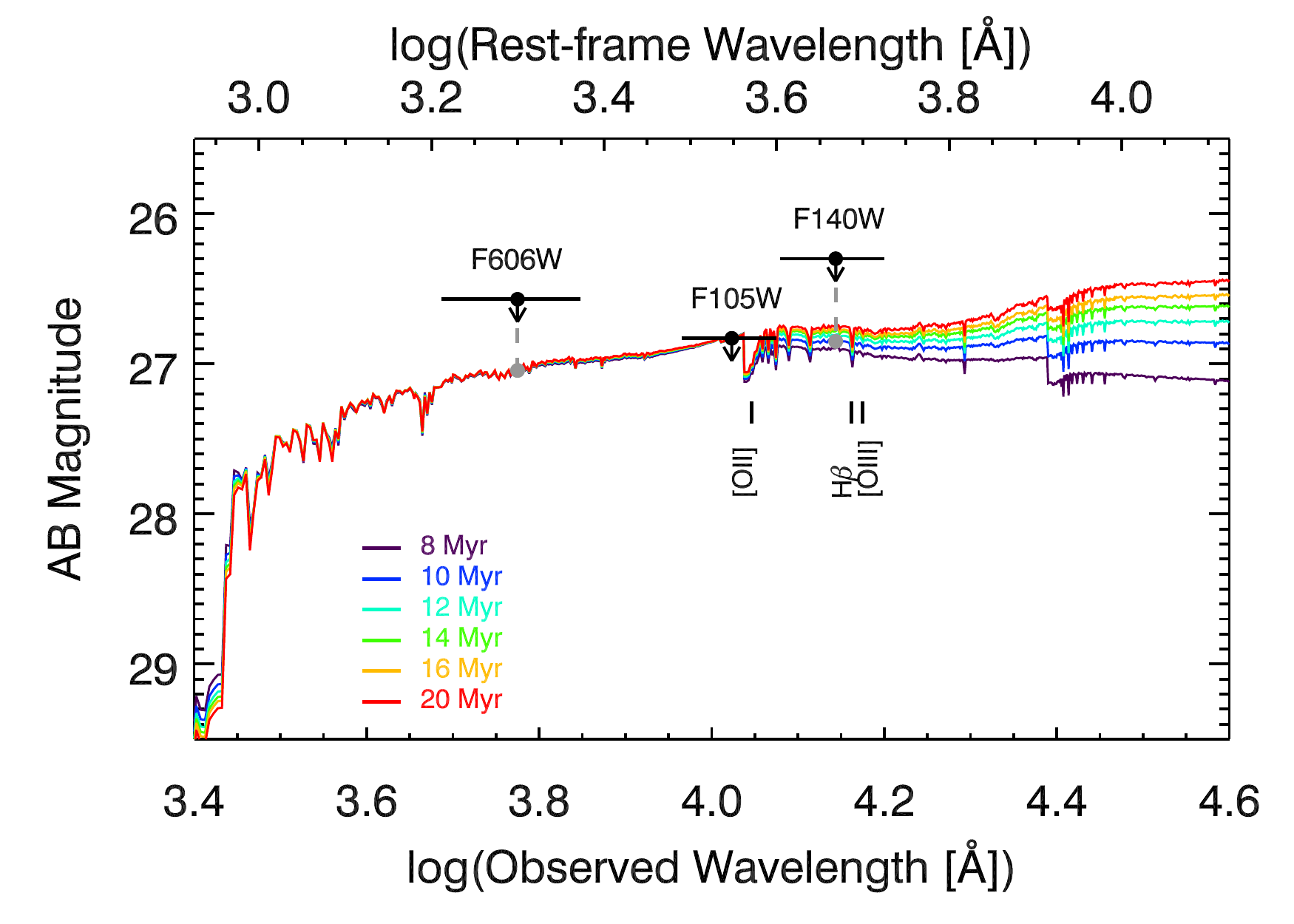}
\caption*{Extended Data Figure 5: Clump continuum flux upper limits. The observed flux upper
  limits estimated from simulations and GALFIT modelling
  in the three bands are shown as black filled circles. The black horizontal
  lines indicate the bandpass width of each filter. Colored curves
  represent reddened Starburst99 stellar population synthesis models$^{43}$
  with different ages (from 8 to 20 Myr), normalized to the most
  stringent upper limit (F105W band). The corresponding upper limits
  in F140W and F606W obtained considering a spectrum with an age
  $\sim$ 10 Myr  are shown as grey filled circles.}
\label{fig:sb99}
\end{figure}

\clearpage

\begin{figure}[h!]
\centering
\includegraphics[width=\textwidth]{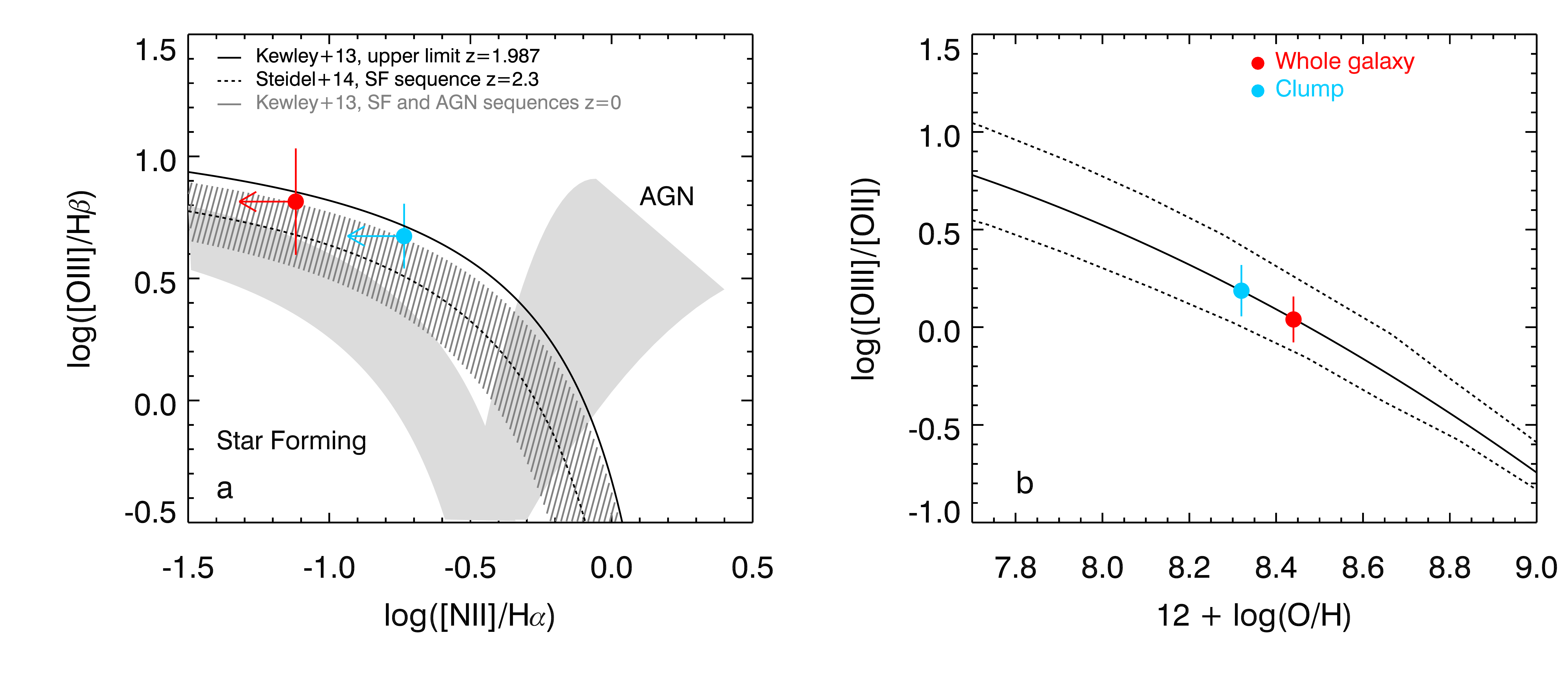}
\caption*{Extended Data Figure 6: Emission line diagnostics.
The BPT diagram$^{54}$ (\textbf{a}) shows that the emission line
ratios of the whole galaxy and
of the clump (red and light blue points with s. d. error bars) are consistent with being powered by
star formation. The [NII]
upper limit and \ha ~emission of the whole galaxy are measured from the Subaru/MOIRCS longslit
spectroscopy follow-up and the \nii/\ha ~upper limit for the clump
is computed assuming the
\nii ~of the whole galaxy. The metallicities of the whole galaxy 
and that of the clump have been determined from the [OIII]/[OII]
ratio$^{64}$ (\textbf{b}).}
\label{fig:bpt}
\end{figure}

\clearpage

\begin{figure}[h!]
\centering
\includegraphics[width=\textwidth]{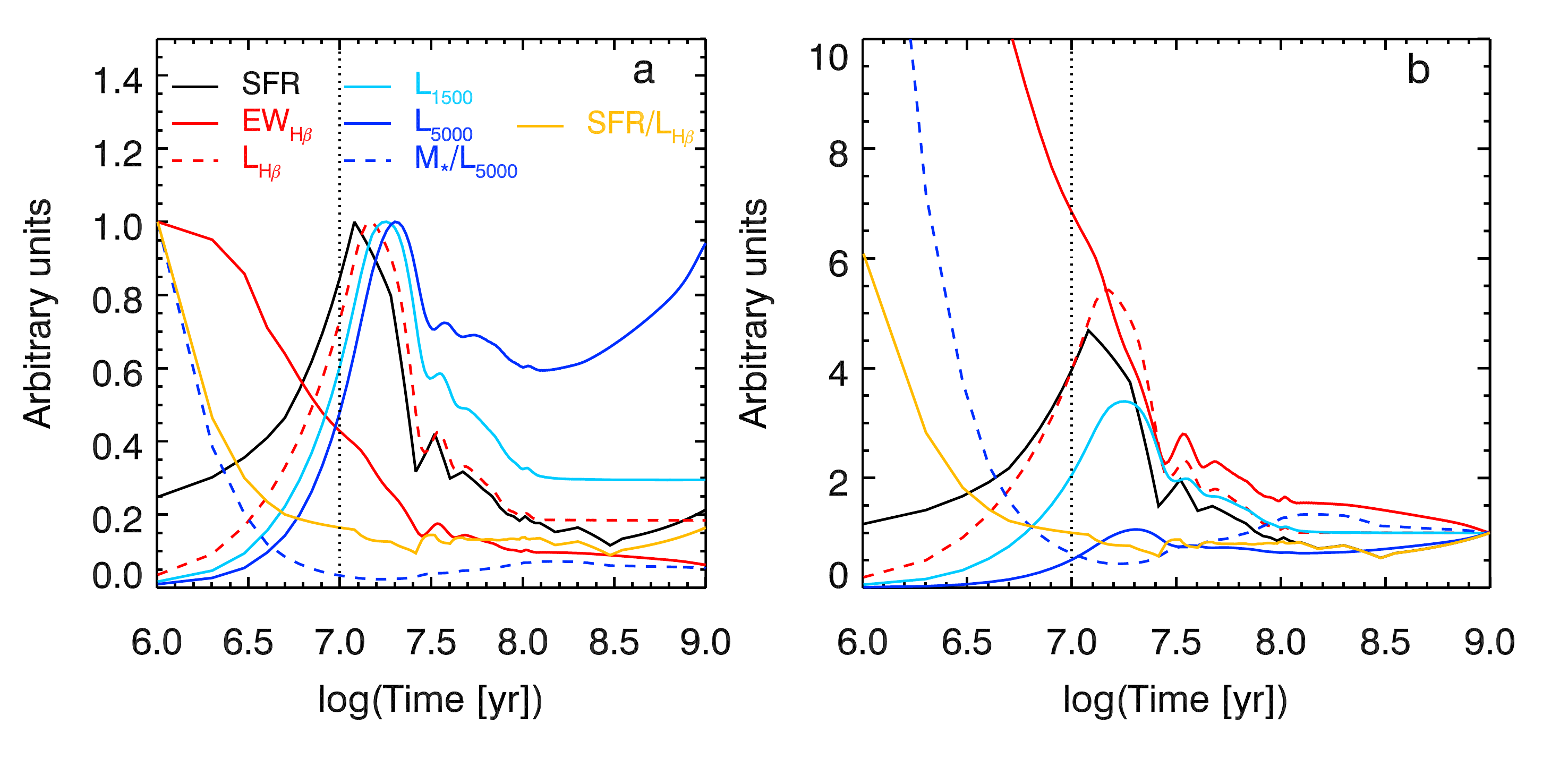}
\caption*{Extended Data Figure 7: Time evolution of physical quantities based on the clump SFR(t)  from our simulations. In panel \textbf{a} the peak of all the
  curves is normalized to 1 to highlight the time delay occuring
  between the peak of the SFR and of the luminosities
  $\mathrm{L_{H\beta}}$, $\mathrm{L_{1500\AA}}$, $\mathrm{L_{5000\AA}}$,
  whereas in panel \textbf{b} they are normalized to 1 at $\mathrm{t =
    1\,Gyr}$ to stress the relative intensity of the observables at
 the peak and later phases. The vertical black dotted line indicates the upper limit on the age of the clump (t = 10 Myr). The units of the plotted quantities are: SFR ($\mathrm{M_{\odot}\,yr^{-1}}$), $\mathrm{EW_{H\beta}}$ ($\mathrm{\AA}$), $\mathrm{L_{H\beta}}$, $\mathrm{L_{1500\AA}}$, $\mathrm{L_{5000\AA}}$ (erg\, s$^{-1}$), $\mathrm{M_{*}/L_{5000\AA}}$ ($\mathrm{M_{\odot}\, erg^{-1}\, s}$), $\mathrm{SFR/L_{H\beta}}$ ($\mathrm{M_{\odot}\, yr^{-1}\, erg^{-1}\, s}$).}
\label{fig:summary}
\end{figure}

\clearpage

\textbf{Tables}
\\
\\
\begin{table}[h!]
\small
\centering
\caption*{Extended Data Table 1: Properties of the galaxy and the clump.
\\
\scriptsize
Notes: $^a$ The effective radius of the galaxy is the average of the $\mathrm{R_e}$ obtained from a single S\'ersic profile fit in the F140W, F105W and F606W imaging. $^b$ The gas mass of the galaxy has been determined, given its SFR, as: $\mathrm{M_{gas} = 9.18 + 0.83\, log(SFR)}^{29}$. $^c$ The observed flux of the F140W, F105W and F606W direct images has been determined with GALFIT. We associated a standard uncertainty of 5\%.}
\label{tab:properties}
\begin{tabular}{p{5cm} c c}
\toprule
\midrule
                      &      Galaxy (ID568)                  & Clump (Vyc1)    \\
\midrule
\textbf{Right ascension} [h m s]  & 14:49:12.578 & 14:49:12.575\\
\textbf{Declination} [$^{\circ}$ ' "]   &   +8:56:19.42 & +8:56:19.62\\
$\mathrm{\mathbf{R_e}}$ [kpc]   &  $2.8\pm 0.4^a$ &  $< 0.5$  \\
\textbf{SFR} [$\mathrm{M_{\odot}/yr}$] & $77 \pm 9$ & $32 \pm 6$ \\
\textbf{log(M$_{\star}/\mathrm{M_{\odot}}$)} & $\mathrm{10.3^{+0.2}_{-0.3}}$ & $\lesssim 8.5$  \\
\textbf{log(M$_{\mathrm{gas}}/\mathrm{M_{\odot}}$)} & $10.7 \pm 0.2^b$   & $\lesssim 9.4$ \\
\textbf{Z} [$\mathrm{Z_{\odot}}$] & $0.6 \pm 0.2$ &  $0.4 \pm 0.2$ \\
$\mathrm{\mathbf{F^{obs}_{[OIII]}}}$ [$\mathrm{10^{-17}erg\, s^{-1}cm^{-2}}$] & $10.4 \pm 0.7$ & $4.3 \pm 0.2$ \\
$\mathrm{\mathbf{F^{obs}_{H\beta}}}$ [$\mathrm{10^{-17}erg\, s^{-1}cm^{-2}}$] & $1.5 \pm 0.8$ & $0.9 \pm 0.3$ \\
$\mathrm{\mathbf{F^{obs}_{[OII]}}}$ [$\mathrm{10^{-17}erg\, s^{-1}cm^{-2}}$] & $6.5\pm 1.7$ & $1.9\pm 0.6$ \\
$\mathrm{\mathbf{F^{obs}_{F140W}}}$ [$\mathrm{10^{-20}erg\, s^{-1}cm^{-2}\, \AA^{-1}}$]    & $67.5  \pm 3.4^c$   &   $<\, 1.1$ \\
$\mathrm{\mathbf{F^{obs}_{F105W}}}$ [$\mathrm{10^{-20}erg\, s^{-1}cm^{-2}\, \AA^{-1}}$]    & $89.2 \pm 4.6^c$  &   $<\, 1.8$  \\
$\mathrm{\mathbf{F^{obs}_{F606W}}}$ [$\mathrm{10^{-20}erg\, s^{-1}cm^{-2}\, \AA^{-1}}$]    & $212.3\pm 10.6^c$   &  $<\, 4.5$ \\
\bottomrule
\\
\end{tabular}
\end{table}

\clearpage

\begin{table}[h!]
\centering
\caption*{Extended Data Table 2: \textit{HST}/WFC3 and Subaru/MOIRCS observations.}
\label{tab:observations}
\begin{tabular}{p{2.5cm} c c c c }
\toprule
\midrule
Instrument             & Date & Time                    & Time      \\
                              &          & (direct imaging)   & (spectroscopy)   \\
                             &          &  (hr)                     &  (hr)        \\
\midrule
\textit{HST}/WFC3              &  2010, 6$^{\mathrm{th}}$ June                & 0.3 (F140W) & 2.7  \\
\textit{HST}/WFC3              &  2010, 25$^{\mathrm{th}}$ June, 1$^{\mathrm{st}}$ July   & 0.6 (F140W) & 7     \\
\textit{HST}/WFC3              &  2010, 9$^{\mathrm{th}}$ July                 & 0.3 (F140W) & 2.7  \\
\textit{HST}/WFC3              &  2013, 20$^{\mathrm{th}}$ May               &  3.3 (F105W)  &  -    \\ 
\textit{HST}/WFC3               &  2013, 20$^{\mathrm{th}}$ May              &  0.3  (F606W)  &  -    \\
Subaru/MOIRCS       & 2013, 7$^{\mathrm{th}}$ - 9$^{\mathrm{th}}$ April                &   -  &  7.3 \\  
\bottomrule
\end{tabular}
\end{table}

\end{document}